\documentclass[aip,cha,reprint]{revtex4-2}

\usepackage{newtxtext}
\usepackage{amsmath}
\usepackage{amssymb}
\usepackage[T1]{fontenc}
\usepackage{graphicx}
\usepackage{dcolumn}
\usepackage{mathtools}
\usepackage{relsize}
\usepackage{subcaption}
\usepackage{braket}
\usepackage{booktabs}
\usepackage{multirow}
\usepackage{xcolor}
\usepackage{etoolbox}
\usepackage{algorithm}
\usepackage{algpseudocode}
\usepackage{caption} 
\usepackage{bm} % Need to be loaded at the end

\newcommand{\argmin}[1]{\mathop{\mathrm{arg\,min}}\limits_{#1}}
\newcommand{\closedots}{\hbox to 0.8em{.\hss.\hss.}}


\begin{document}

\title{Encoding electronic ground-state information with variational even-tempered basis sets}

\author{Weishi Wang}
\email{weishi.wang.gr@dartmouth.edu}
\affiliation{Department of Physics and Astronomy, Dartmouth College, Hanover, NH 03755, USA}

\author{Casey Dowdle}
\affiliation{Department of Mathematics, Dartmouth College, Hanover, NH 03755, USA}

\author{James D. Whitfield}
\affiliation{Department of Physics and Astronomy, Dartmouth College, Hanover, NH 03755, USA}
\affiliation{AWS Center for Quantum Computing, Pasadena, CA 91125, USA}

\begin{abstract}
We propose a system-oriented basis-set design based on even-tempered basis functions to variationally encode electronic ground-state information into molecular orbitals. First, we introduce a reduced formalism of concentric even-tempered orbitals that achieves hydrogen energy accuracy on par with the conventional formalism, with lower optimization cost and improved scalability. Second, we propose a symmetry-adapted, even-tempered formalism specifically designed for molecular systems. It requires only primitive S-subshell Gaussian-type orbitals and uses two parameters to characterize all exponent coefficients. In the case of the diatomic hydrogen molecule, the basis set generated by this formalism produces a dissociation curve more consistent with cc-pV5Z than cc-pVTZ at the size of aug-cc-pVDZ. Finally, we test our even-tempered formalism against several types of tetra-atomic hydrogen molecules for ground-state computation and point out its current limitations and potential improvements.
\end{abstract}

\maketitle

\section{Introduction}\label{sec:1}

Orbitals are essential building blocks for the many-electron wavefunctions~\cite{szabo2012modern, helgaker2013molecular}. In non-interacting systems, the eigen wavefunctions are precisely described by the Slater determinants of spin-orbitals that are the eigenfunctions of the separated one-body problems. For interacting systems, initializing the many-electron reference states with proper orbitals can also improve the accuracy or compactness of the downstream multi-configurational ground-state ansatzes.

The strategies for constructing optimal orbitals can be divided into two main types. Fitting the target orbitals (implicitly determined by an ansatz) through the linear combinations of a given basis set (e.g., orbital rotations) and directly engineering the basis set form such that it can efficiently span a function space close to where those orbitals reside. Many optimization procedures, such as the Hartree--Fock methods~\cite{slater1930note, fock1930naherungsmethode, echenique2007mathematical} and orbital localization methods~\cite{edmiston1963localized, magnasco1967uniform, li2014localization}, fall into the first type, whereas basis-set designs belong to the second. Atomic basis sets~\cite{jensen2013atomic} built on contracted Gaussian-type orbitals (GTOs)~\cite{boys1950electronic} have been a successful design for molecular electronic structure. This is due to their efficient fitting of atomic orbitals despite missing the center cusps present in Slater-type orbitals (STOs)~\cite{slater1930atomic, hehre1969self}. In fact, many reusable atomic Gaussian basis sets have been proposed for various systems and electronic structure methods~\cite{jensen2013atomic, schuchardt2007basis}. Nevertheless, there still lacks a unified framework to construct system-oriented basis sets without empirical contractions or parameterizations.

Before the rapid development of tabulated atomic Gaussian basis sets, one of the approaches to construct system-oriented basis sets was by directly optimizing GTOs with respect to the system's ground-state energy~\cite{reeves1963use}. To reduce the number of free exponent coefficients during the optimization, Reeves et al. proposed the ideas behind even-tempered basis sets~\cite{reeves1963use2, raffenetti1973even, bardo1973even, bardo1973even2, raffenetti1973even2, bardo1974even}. Restricted to S-subshell orbitals, even-tempered basis functions are defined as a sequence of concentric Gaussian functions
\begin{equation}\label{eq:etb1_1}
    \phi_{mn}\left(\bm{r}\,\vrule\,\alpha_m\right) \equiv 
        \mathcal{N}\!\left(\alpha_m\right)\,{\rm{exp}}\left(-\alpha_m\,\lVert\,\bm{r}-\bm{R_n}\,\rVert^{2}_2\right),
\end{equation}
where $\bm{R}_n$ is the center coordinate in the atomic units (a.u.),  $\mathcal{N}\!\left(\alpha_m\right) \!=\! \left(2\alpha_m/\pi\right)^{3/4}$ is the normalization factor, and $\alpha_m$ are the exponent coefficients correlated by a  pair of primitive parameters $\bm{\gamma} \equiv \left(\alpha,\,\beta\right)$:
\begin{equation}\label{eq:etb1_2}
    \alpha_m = \alpha\,\beta^{\,m},\qquad \alpha>0,\;\beta>0,\;m = 0,1,2,\dots
\end{equation}
For consistency, we apply the atomic unit system to all the physical quantities presented in this paper. Particularly, we set the unit of $\alpha$ to be $1/$a.u. (i.e., the inverse of the Bohr radius $a_{\rm B}$) and let $\beta$ remain unitless. We do not explicitly denote the unit of $\alpha$ throughout the rest of this paper for simplicity.

It has been empirically shown that, by optimizing $\bm{\gamma}$ against atomic Hartree--Fock energy, even-tempered basis sets can systematically converge to the complete basis-set (CBS) limit~\cite{kryachko2003generation, bakken2004expansion}. The accuracy of the resulting basis sets does not directly transfer to excited-state computations. However, this issue can be resolved by re-optimizing the basis sets directly with respect to the excited state energy~\cite{glushkov2006excited}. There have been multiple efforts to analytically explain the effectiveness of even-tempered basis sets for atomic systems. In the case of the hydrogen atom, the construction of even-tempered basis sets is justified as a discretization of an integral transformation for hydrogen atomic orbitals~\cite{feller1979systematic, kutzelnigg1994theory, kutzelnigg1996convergence}. Alternatively, they can be seen through the lens of the Gram-Schmidt process applied to Gaussian functions~\cite{cherkes2009spanning}. Progress has also been made on rigorously bounding the error convergence. Specifically, Kutzelnigg~\cite{kutzelnigg1994theory} showed that the ground-state energy error of the hydrogen atom asymptotically scales as $\exp(-d \sqrt{m})$ for a basis set of size $m$, where the positive constant $d$ is dependent on $\bm{\gamma}$. Bachmayr et al. further improved upon this study and established the error bounds on how well even-tempered basis sets approximate Slater-type functions beyond the S subshell~\cite{bachmayr2014error}.

Today, even-tempered basis sets are primarily used to augment atomic Gaussian basis sets for the purpose of capturing diffuse orbitals consistently \cite{morgan2015additional,koput2015ab}. Optimizing even-tempered basis sets directly for molecular systems, though briefly attempted~\cite{bardo1974even}, remains an overlooked direction. As programming paradigms and hardware architectures advance, more sophisticated construction and optimization of basis sets become more feasible through modern scientific software~\cite{schoendorff2021development, wang2023basis, shaw2023basisopt}. Repositioning even-tempered basis sets as a discretization strategy rather than an augmentation measure, we pose the following question: How efficiently can a system-oriented even-tempered basis set encode electronic ground-state information beyond atomic systems? To answer this question, we introduce a simple basis-set construction formalism based on even-tempered GTOs in Equation~(\ref{eq:etb1_1}), which can be directly applied to atomic and molecular systems. We further analyze the accuracy and the stability of even-tempered basis functions with respect to their tunable parameters. We draw implications from these numerical analyses and propose new hierarchical variational optimization procedures tailored for our formalism with improved numerical stability. By combining the construction formalism and optimization strategy, we demonstrate a system-oriented and tabular data-free basis-set design.

The paper is organized as follows. In Sec.~\ref{sec:2}, we first define a reduced formalism of even-tempered basis sets and formalize the concept of variational even-tempered basis sets. Second, we propose a basic one-level optimization strategy that focuses on optimizing the exponent coefficients for reduced even-tempered basis sets in Sec.~\ref{sec:3}. We then apply it to atomic hydrogen to quantify the correlation between $\alpha$ and $\beta$, as well as their impact on the stability of ground-state computation. Next, in Sec.~\ref{sec:4}, we introduce the updated two-level optimization strategy that incorporates the optimization of basis-function centers for molecular systems. We numerically test its performance on various geometric configurations of diatomic and tetra-atomic hydrogen systems. Last, we conclude our paper in Sec.~\ref{sec:5} with discussions about the limitations of our design and potential future directions.

\section{Variational even-tempered basis sets}\label{sec:2}

To study the effectiveness of constructing electronic ground states with system-oriented even-tempered basis sets, we propose a modified formalism to analyze the impact of $\bm{\gamma}$ on subsequent electronic-structure computations. We first define a reduced sequence of even-tempered  
exponent coefficients:
\begin{equation}
    \zeta_m = \alpha\,\beta^{\,m},\qquad \alpha>0,\;\beta>0,\;m = 1,2,\dots
\end{equation}
Compared to the conventional form in Equation~(\ref{eq:etb1_2}), index $m$ in this new form iterates from one instead of zero, which ensures that $\alpha$ and $\beta$ both exist for all possible $\zeta_m$. Correspondingly, one can vary $\beta$ to analyze its effect on the downstream computations while having $\alpha$ fixed at a global value, or vice versa. Such a subtle modification in parameterizing the exponent coefficients paves the way for an improved variational optimization strategy of even-tempered basis sets, which we shall introduce in Sec.~\ref{sec:3}.

Formally, we define a \textbf{reduced even-tempered basis set} as
\begin{equation}\label{eq:ret_basis}
    \mathcal{G}^{\rm r}_{M} \equiv \left\{\phi_{mn}\left(\bm{r}\,\vrule\,\zeta_m\right)\,\vrule\,m=1,\dots,M;\,\,n=1,\dots,N\right\}, 
\end{equation}
where $M$ is the \textbf{basis degree} that determines the multiplicity of concentric even-tempered basis functions, while $N$ specifies the number of basis-function centers $\bm{R_n}$. Therefore, the basis-set size $\vert\mathcal{G}^{\rm r}_M\vert$ is equal to $M\!\times\!N$. At the boundary configuration $\mathcal{G}^{\rm r}_1$, the effects of $\alpha$ and $\beta$ converge, as the optimal value of $\bm{\gamma}$ is degenerate with respect to their combinations. We also note that the basis functions in $\mathcal{G}_{M}$, whether or not having the same centers, are all characterized by $\bm{\gamma}$. This simplification further reduces the cost of basis-set parameter optimization. 

In comparison to $\mathcal{G}^{\rm r}_{M}$, we define the \textbf{non-reduced even-tempered basis set} based on the conventional formalism as 
\begin{equation}\label{eq:oet_basis}
    \mathcal{G}^{\rm o}_{M} \equiv \left\{ \phi_{(m-1)n}\!\left(\bm{r}\,\vrule\,\alpha_{m-1} \right)\,\vrule\, m,\,n \right\},
\end{equation}
where $m$ and $n$ still iterates from $1$ to $M$ and $N$, respectively. Throughout the rest of this paper, we shall use symbol ``$\mathcal{G}_{M}$'' for the statements where both the reduced and non-reduced formalisms apply.

Given a target many-electron system $\mathbb{M}$ (under specific nuclear geometry and electronic spin configuration), let the parameterized even-tempered basis set be $\mathcal{G}_{M}\!\left(\bm{\Theta}\right)$, where $\bm{\Theta}$ is a (vectorized) set of all its tunable parameters that are distinguished by their symbols rather than assigned values. We define a partial ground-truth variational optimization of $\mathcal{G}_{M}\!\left(\bm{\Theta}\right)$ as 
\begin{equation}\label{eq:bs_opt}
    \bm{\tilde{\theta}} = \argmin{\bm{\theta} \subseteq \bm{\Theta}}\hat{\mathcal{E}}\!\left(\mathbb{M}\right)\!\left[\mathcal{G}_{M}\!\left(\bm{\Theta}\right)\right], 
\end{equation}
where $\bm{\tilde{\theta}}$ is the optimized value for a parameter subset $\bm{\theta}$ of $\bm{\Theta}$; $\hat{\mathcal{E}}\!\left(\mathbb{M}\right)$ is ground-state energy ansatz, treated as a functional of $\mathcal{G}_M$, parameterized by $\mathbb{M}$. Moreover, we focus on the optimization of even-tempered basis sets with respect to the Hartree--Fock (HF) energy functional ($\hat{\mathcal{E}} \coloneqq \hat{\mathcal{E}}_{\rm HF}$). In this way, the accuracy of the ground-state energy approximation directly reflects the performance of the basis-set design, since the HF approximation only relies on orbital rotation. For the gradient-based implementation of the optimization procedure in Equation~(\ref{eq:bs_opt}), we refer the reader to the technical literature of Quiqbox~\cite{wang2023basis}, an open-source basis-set software library that we used to construct or optimize basis sets. Finally, we define a \textbf{variational even-tempered basis set} with respect to the optimized parameter subset $\bm{\theta}$ as
\begin{equation}
    \tilde{\mathcal{G}}_{M}\!\left(\bm{\Theta}\setminus{ \bm{\theta}}\right)\equiv\mathcal{G}_{M}\!\left(\bm{\Theta}\,\vrule_{\,\bm{\theta}=\bm{\tilde{\theta}}}\right).
\end{equation}

Based on Equations~(\ref{eq:ret_basis}) and (\ref{eq:oet_basis}), we divide the adjustable coefficients of $\mathcal{G}_{M}$ into two levels. On the first level, $\{\zeta_m\,\vrule\,m\}$ (or $\{\alpha_m\,\vrule\,m\}$) affects how the basis functions centered at a given $\bm{R_n}$ span a one-body Hilbert space~\cite{cherkes2009spanning}, which is a subspace of $L^2(\mathbb{R}^3)$. On the second level, $\{\bm{R_n}\,\vrule\,n\}$ further determines the placements of the basis functions in real space $\mathbb{R}^3$. Specifically, at the limit of $N\!=\!1$ with a fixed $\bm{R_1}$, $\{\zeta_m\,\vrule\,m\}$ affects the ``resolution'' of $\mathcal{G}^{\rm r}_{M}$; at the limit of $M\!=\!1$ with a fixed $\{\zeta_m\,\vrule\,m\}$, $\{\bm{R_n}\,\vrule\,n\}$ determines the ``volume'' of $\mathcal{G}^{\rm r}_{M}$. However, when $M$ and $N$ are both larger than one, the effects from these two levels are not necessarily orthogonal. Hence, the key to generating system-oriented $\mathcal{G}_{M}$ lies in efficiently optimizing the primitive parameters (e.g., $\bm{\gamma}$) that control these two levels of basis-set coefficients.

\section{One-level variational optimization}\label{sec:3}

In this section, we only consider the effect of $\bm{\gamma}$. Therefore, we focus on atomic systems for which even-tempered basis functions were originally considered. In this context, the potential benefit of placing the basis functions on locations other than the nucleus can be eliminated. Furthermore, we fix the values of $\alpha$ to preset global values, and only optimize $\beta$ in the reduced formalism. In addition to applying the parameter optimization routine from Quiqbox, we propose a bootstrap strategy that iteratively uses the results of $\tilde{\mathcal{G}}^{\rm r}_{M-1}$ to jump-start the optimization for $\tilde{\mathcal{G}}^{\rm r}_M$. The pseudo-code of this strategy is presented in Algorithm \ref{alg:mop1}.

\begin{figure}[htb] % To make algorithm block float
\begin{algorithm}[H] % MUST use [H] to avoid compilation errors
    \caption{$\beta$-adapted even-tempered basis optimization}\label{alg:mop1} % Only optimize \beta
    \begin{algorithmic}[1]
        \renewcommand{\baselinestretch}{1.4}\selectfont
        \State $\tilde{\beta} \coloneqq 1$
        \State Initialize $\hat{\mathcal{E}}\!\left(\mathbb{M}\right)$, $\alpha$, $M_{\rm max}$
        \For {$M$ in $\left(1,\dots,M_{\rm max}\right)$}
            \State Initialize $\mathcal{G}^{\rm r}_{M}\!\left(\{\bm{\gamma}\}\right)$ with $\bm{\gamma}\coloneqq(\alpha,\,\tilde{\beta})$
            \State $E_{\rm HF} \coloneqq \hat{\mathcal{E}}\left[\mathcal{G}^{\rm r}_{M}\right]$, $i\coloneq0$, $\Delta\coloneqq\inf$
            \While {$i < i_{\rm max}$ and $\vrule\,\Delta\,\vrule > \delta$}
                \State ${
                    \setlength{\jot}{0.0em}
                    \begin{aligned}
                        \text{7.1)}\hspace{-2mm} & &i &\coloneqq\,i + 1 \\
                        \text{7.2)}\hspace{-2mm} & &\tilde{\beta} &\coloneqq\,{\rm{optimizer}}\!\left[\hat{\mathcal{E}},\,\partial \hat{\mathcal{E}}/\partial \beta,\,\mathcal{G}^{\rm r}_{M}\right] \\
                        \text{7.3)}\hspace{-2mm} & &\bm{\gamma} &\coloneqq\,(\alpha,\,\tilde{\beta}) \\
                        \text{7.4)}\hspace{-2mm} & &\mathcal{G}^{\rm r}_{M} &\coloneqq\, \mathcal{G}^{\rm r}_{M}\!\left(\{\bm{\gamma}\}\right) \\ 
                        \text{7.5)}\hspace{-2mm} & &\Delta &\coloneqq\,E_{\rm HF} - \hat{\mathcal{E}}\left[\mathcal{G}^{\rm r}_{M}\right] \\
                        \text{7.6)}\hspace{-2mm} & &E_{\rm HF} &\coloneqq\,E_{\rm HF} - \Delta
                    \end{aligned}
                }$
            \EndWhile
            \State Store $\left(M,\,\tilde{\beta},\,E_{\rm HF}\right)$
        \EndFor
    \end{algorithmic}
\end{algorithm}
\end{figure}

The advantage of Algorithm \ref{alg:mop1}'s bootstrap initialization is that, by construction, the initial guess of $\tilde{\mathcal{G}}^{\rm r}_{M\!+\!1}$ is a proper superset of $\tilde{\mathcal{G}}^{\rm r}_M$:
\begin{equation}\label{eq:basis_ext}
    \mathcal{G}^{\rm r}_{M\!+\!1}\,\vrule_{\,i=0}
    \!=\!\tilde{\mathcal{G}}^{\rm r}_M \cup \left\{\phi_{(M\!+\!1)n}\!\left(\bm{r};\bm{\gamma}\,\vrule_{\,\beta=\tilde{\beta}\left(\alpha;\,M\right)},\bm{R_n}\right)\,\vrule\,n\right\}. 
\end{equation}
Hence, by the variational principle, it imposes the ground-state energy bound
\begin{equation}
\hat{\mathcal{E}}\!\left(\mathbb{M}\right)\!\left[\mathcal{G}^{\rm r}_{M\!+\!1}\,\vrule_{\,i=0}\right] \leq \hat{\mathcal{E}}\!\left(\mathbb{M}\right)\!\left[\tilde{\mathcal{G}}^{\rm r}_M\right]
\end{equation}
such that extending a reduced variational even-tempered basis from degree $M$ to $M\!+\!1$ never worsens its performance. 

\subsection{Ground-state energy of hydrogen atom}

For the probing system, we choose the simplest atom: hydrogen. As an analytically solvable system, it has been tested in several literature~\cite{schmidt1979effective, klopper1986gaussian, kutzelnigg1994theory, bakken2004expansion} to study the scaling performance of even-tempered basis sets. Moreover, since the hydrogen atom is free of electron-electron interaction, its HF energy $E_{\rm HF}$ converges to the true ground-state energy $E_{\rm GT}$ ($-0.5$ Ha) at the CBS limit. 

\begin{figure}[h]
    \centering
    \includegraphics[width=\linewidth]{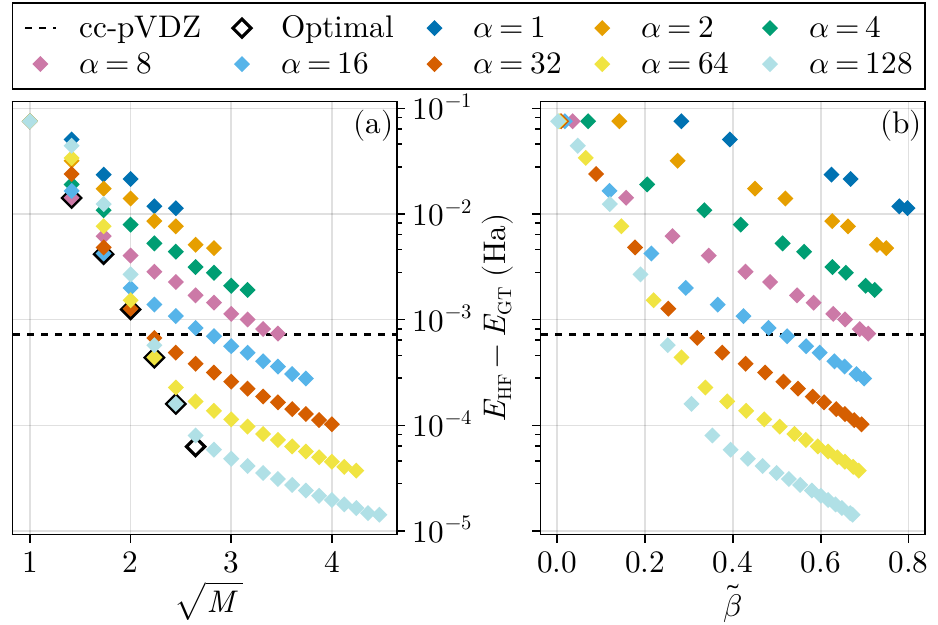}
    \caption{The atomic hydrogen ground-state energy error of the reduced variational even-tempered basis set $\tilde{\mathcal{G}}^{\rm r}_M$ (the solid colored diamonds) with respect to $M$ (equivalent to basis set size in this case) and $\alpha$. (a) shows the logarithmic errors, where the open diamonds with black strokes represent the conventionally optimized even-tempered basis set $\tilde{\mathcal{G}}^{\rm o}_M$ at different $M$ (incremented by one). (b) shows the relation between the error and optimized parameter $\tilde{\beta}$ for each configuration of $\tilde{\mathcal{G}}^{\rm r}_M$. For each color group, the data points in (b) follow the same correspondence to $M$ as the ones in (a), respectively.}
    \label{fig:H_energy}
\end{figure}

We use the optimized values of $\bm{\gamma}$ from reference~\cite{bakken2004expansion} to construct the conventional counterpart of $\tilde{\mathcal{G}}^{\rm r}_M$, $\tilde{\mathcal{G}}^{\rm o}_M$, due to its high-precision numerical computation (70-digit precision according to the authors). To verify the results from this reference, we recomputed the ground-state energies (from $M\!=\!2$ to $M\!=\!10$) using its parameter settings in a 128-bit floating number system and obtained a mean absolute error (MAE) $\sim\!2\times 10^{-10}$ with respect to its reported data. We generated a series of $\tilde{\mathcal{G}}^{\rm r}_M$ from the optimized $\tilde{\beta}$ using Algorithm~\ref{alg:mop1} under multiple values of $\alpha$ and $M$. The comparison of the ground-state energies between $\tilde{\mathcal{G}}^{\rm r}_M$ and $\tilde{\mathcal{G}}^{\rm o}_M$ is plotted in FIG.~\ref{fig:H_energy}. To check the values of the representative energies and the corresponding $\tilde{\beta}$, we refer the reader to TABLE~\ref{tab:rhf-hydrogen} in Appx.~\ref{app:1}.

As provided by previous studies~\cite{klopper1986gaussian, kutzelnigg1994theory}, the asymptotic energy error of $\tilde{\mathcal{G}}^{\rm o}_M$ follows an exponential decay with respect to the square root of its basis-set size:
\begin{equation}\label{eq:energy_decay}
    \ln \left(\vert E_{\rm HF} - E_{\rm GT} \vert \right) \propto \sqrt{\vert\tilde{\mathcal{G}}^{\rm o}_M\vert}.
\end{equation}
This relation was verified by works~\cite{bakken2004expansion, bachmayr2014error} and is reproduced again in FIG.~\ref{fig:H_energy}(a) by the open diamonds (data points marked as ``Optimal'' for $\tilde{\mathcal{G}}^{\rm o}_M$), as $M$ is equal to the basis-set size $\vert\mathcal{G}_M\vert$ for atomic systems. Unsurprisingly, the energy errors of $\tilde{\mathcal{G}}^{\rm o}_M$ form lower bounds for $\tilde{\mathcal{G}}^{\rm r}_M$. Additionally, we found that there exists a numerically consistent match between the errors of $\tilde{\mathcal{G}}^{\rm o}_M$ and $\tilde{\mathcal{G}}^{\rm r}_M$ with exponentially increased $\alpha$ in FIG.~\ref{fig:H_energy}(a): a sequence of colored diamonds in which each one is enclosed by an open diamond. 

\begin{figure}[h]
    \centering
    \includegraphics[width=\linewidth]{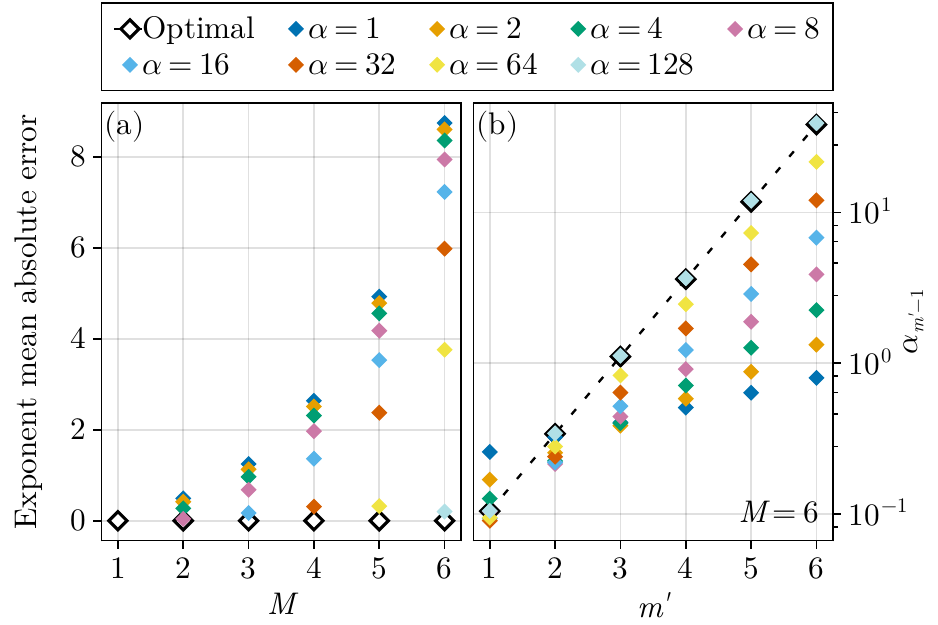}
    \caption{The errors of the exponents $\left\{\zeta_m\vert m\right\}$ from $\tilde{\mathcal{G}}^{\rm r}_M$ with respect to the optimal exponents $\left\{\alpha_{m'}\vert m'\right\}$ from $\tilde{\mathcal{G}}^{\rm o}_M$. 
    In (a), each colored diamond represents the mean absolute error of  $\left\{\zeta_m\vert m\right\}$ (rearranged in ascending order to match $\left\{\alpha_{m'}\vert m'\right\}$) from $\tilde{\mathcal{G}}^{\rm r}_M(\alpha)$. (b) shows that for $M\!=\!6$, the rearranged exponents of $\tilde{\mathcal{G}}^{\rm r}_M(\alpha)$ converge to their respective optimal values (the open diamonds chained by a dotted line) as $\alpha$ increases exponentially from $1$ to $128$.}
    \label{fig:H_shape}
\end{figure}

To verify whether this matching pattern is also reflected on the exponent coefficients, we computed the mean absolute errors (MAE) of $\left\{\zeta_m\,\vert\,m=1,\dots,M\right\}$ for different $\tilde{\mathcal{G}}^{\rm r}_M(\{\alpha\})$ against $\left\{\alpha_{m'-1}\,\vert\,m'=1,\dots,M\right\}$ from the respective $\tilde{\mathcal{G}}^{\rm o}_M$. For each computed MAE, $\{\zeta_m\vert m\}$ is rearranged to form the same ascending order as $\{\alpha_{m'}\vert m'\}$. The results are shown in FIG.~\ref{fig:H_shape}(a), where each colored diamond represents the MAE for one configuration $(M,\,\alpha)$. As $M$ increases, the optimal value of $\alpha$ grows exponentially. More specifically, $\tilde{\mathcal{G}}^{\rm o}_M$ can be well approximated by $\tilde{\mathcal{G}}^{\rm r}_M(\alpha)$ through following mapping:

\begin{equation}\label{eq:et_map}
    \tilde{\mathcal{G}}^{\rm o}_M \to \tilde{\mathcal{G}}^{\rm r}_M\left(\left\{\alpha= 2^{M+1}\right\}\right) \quad{\rm for}\quad M\geq2.
\end{equation}
This mapping relation indicates that the optimal values of $\alpha$ and $\beta$ are inherently coupled through $M$. Aside from relation (\ref{eq:energy_decay}), we discovered that the scaling of the energy error with respect to $\tilde{\beta}$ also transits to an exponential decay after the value of $\tilde{\beta}$ surpasses the optimal value for a given $\alpha$. This relation is shown in FIG.~\ref{fig:H_energy}(b). To examine how $\alpha$ affects the optimization of $\beta$, we plot the rearranged $\{\zeta_m\vert m\}$ with respect to different values of $\alpha$ in the case of $M\!=\!6$ in FIG.~\ref{fig:H_shape}(b). 

\begin{figure}[h]
    \centering
    \includegraphics[width=\linewidth]{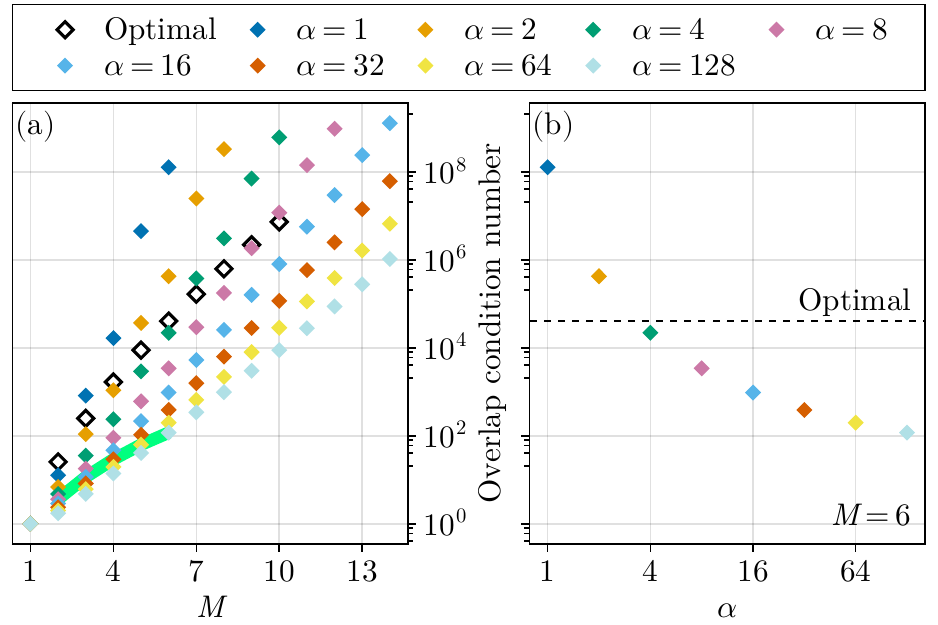}
    \caption{The scaling of the overlap-matrix ($\bm{S}$) condition number for various even-tempered basis set configurations tested against the hydrogen atom. In (a), the colored diamonds represent the results for $\tilde{\mathcal{G}}^{\rm r}_M(\alpha)$, and the open diamonds (marked as ``Optimal'') represent the results for $\tilde{\mathcal{G}}^{\rm o}_M$. The light green stroke passes through the data points of $\tilde{\mathcal{G}}^{\rm r}_M(\{\alpha \!=\! 2^{M+1}\})$. (b) shows how the $\bm{S}$ condition number of $\tilde{\mathcal{G}}^{\rm r}_6(\alpha)$ is suppressed as $\alpha$ increases, where the dashed horizontal line represents the condition number for $\tilde{\mathcal{G}}^{\rm o}_{6}$.}
    \label{fig:H_cond}
\end{figure}

A potential challenge for applying even-tempered basis sets, which has been seldom discussed in the previous literature, is the numerical sensitivity of the corresponding overlap matrix. In general, the overlap matrix of a basis set becomes ill-conditioned when there are near-linearly dependent basis functions. For Gaussian basis sets, this issue can occur as more concentric (or near concentric but diffuse) GTOs with the same angular momentum are added. An ill-conditioned overlap matrix can lead to numerical instability in solving the generalized eigenvalue problem for the free-fermion (or mean-field approximated) eigenstates: 
\begin{equation}
    \bm{H}\bm{C} = \bm{S}\bm{C}\bm{\varepsilon},
\end{equation}
where $\bm{C}$ is the orbital coefficient matrix, $\bm{\varepsilon}$ is the orbital energies as a diagonal matrix, and $\bm{H}$ is the one-body Hamiltonian (or the Fock matrix) discretized by a basis set with its overlap matrix $\bm{S}$. This issue typically can be mitigated by specific orthonormalization strategies~\cite{lowdin1970nonorthogonality, lehtola2019curing} that produce a smaller basis set at the cost of truncating the function space that the original basis sets span. However, within the scope of this paper, which focuses solely on the stability and accuracy of the basis set itself, we refrained from applying these techniques during the HF computation. In the case of the hydrogen atom, we plot the scaling of the overlap-matrix condition numbers for $\tilde{\mathcal{G}}^{\rm o}_M$ and $\tilde{\mathcal{G}}^{\rm r}_M$ in FIG.~\ref{fig:H_cond}, both of which scale exponentially with respect to $M$. In contrast, for $\tilde{\mathcal{G}}^{\rm r}_M(\{\alpha \!=\! 2^{M+1}\})$, which is highlighted by the green stroke, its overlap-matrix condition number scales sub-exponentially. This indicates that applying the mapping of Equation~(\ref{eq:et_map}) can provide a better numerical stability. For a given basis set size, one can still increase $\alpha$ to estimate an asymptotic lower bound of the condition for $\tilde{\mathcal{G}}^{\rm r}_M$, as shown in FIG.~\ref{fig:H_cond}(b).

Although the $\tilde{\mathcal{G}}^{\rm r}_M$ is generated with respect to HF energy, in the case of atomic hydrogen, the energy error directly reflects the capability of using even-tempered Gaussian functions to approximate the ground-state atomic orbital. Hence, Equation~(\ref{eq:et_map}) indicates that one may simplify such orbital fitting procedure into the optimization of a single parameter ($\beta$). This simplified optimization strategy is useful for generalizing approximating arbitrary spherical (S-subshell) orbitals with even-tempered Gaussian functions. 

\section{Two-level variational optimization}\label{sec:4}

In the last section, we quantify the internal correlation between $\alpha$ and $\beta$ in the case of atomic hydrogen and propose a one-level optimization strategy (for $\beta$) in Algorithm~\ref{alg:mop1}. For molecular systems, we need to consider the optimization of not only the exponent parameters (first-level), but also the centers of the even-tempered basis functions (second-level). This is because the distortion of the specific geometry of the centers can help incomplete basis sets fit the target molecular orbitals~\cite{frost1967floating, huber1979geometry, hurley1988computation, wang2023basis}.

To control the shifting (floating) of the basis-function centers, we use implicit parameters that correlate the center coordinates while preserving a desirable symmetry based on the target system. Formally, we define the \textbf{correlated basis centers} with respect to a molecular system $\mathbb{M}$ by
\begin{equation}\label{eq:nc}
    \mathcal{C}_{N}\!\left(\bm{\nu}\right) \equiv \left\{\bm{R_n}\!\left(\bm{\nu}\right)\,\vrule\,n=1,\dots,N\right\},
\end{equation}
where each $\bm{R_n}(\bm{\nu})$ set to be the basis-function center of $\phi_{mn}$ for all $m$, parameterized by $\bm{\nu}$ such that the geometric symmetry formed by $\bm{R_n}$ is invariant under the optimization of $\bm{\nu}$. Furthermore, the implicit parameterization of $\bm{R_n}$ reduces the dimension of the basis-set parameter space compared to the direct optimization of all the coordinate components. For instance, in homonuclear diatomic systems, $\bm{\nu}$ can contain only a single parameter: the distance between the two basis-function centers.

\begin{figure}[htbp]
    \centering
    \includegraphics[width=\linewidth]{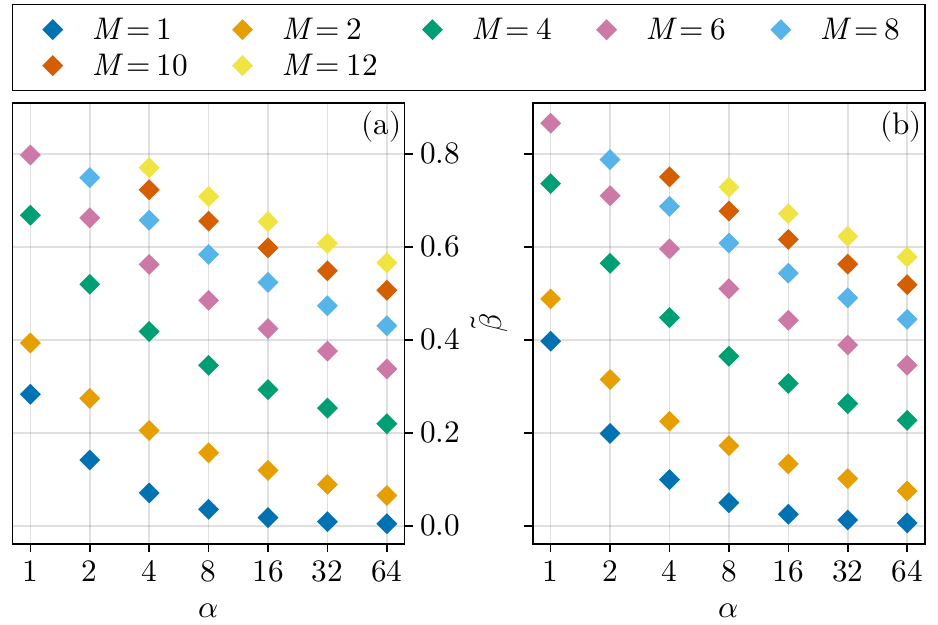}
    \caption{The numerically-stable values of $(\alpha,\,\tilde{\beta})$ with respect to basis degree $M$ of the even-tempered basis sets for: (a) the atomic hydrogen; (b) the diatomic hydrogen (H$_{2}$) molecule at the bond length of $1.4$ a.u.}
    \label{fig:H1H2_param}
\end{figure}

Before incorporating the second-level optimization of basis-function centers, we shall first test the transferability of the one-level optimization strategy (Algorithm~\ref{alg:mop1}) alone for molecular systems. Specifically, we applied it to the diatomic hydrogen (H$_2$) at a bond length of $1.4$ a.u. The results of the $\alpha$--$\tilde{\beta}$ relation compared to the atomic hydrogen are shown in FIG.~\ref{fig:H1H2_param}.

Comparing FIG.~\ref{fig:H1H2_param}(a) and FIG.~\ref{fig:H1H2_param}(b), we can see that despite being qualitatively similar, the quantitative relation between $\tilde{\beta}$ and $\alpha$ for H$_2$ is not the same as that for atomic hydrogen. Subsequently, Equation~(\ref{eq:et_map}) no longer applies. Nevertheless, this divergence is not surprising. In the case of atomic hydrogen, the error of $E_{\rm HF}$ directly represents the discretization error of the basis set for the ground-truth (Slater-type) atomic orbital; whereas for molecular systems, the generated molecular orbitals carry the errors from both the basis discretization and HF approximation of the electron-electron interactions. Admittedly, it is possible to numerically re-derive a new fitting function for $\alpha$. Such an effort would not be an efficient solution when the interpolated coefficients are determined system by system. Therefore, for molecular systems, we instead propose an optimization strategy that imposes an iterative tuning procedure of $\alpha$ on top of the simultaneous optimization of $\tilde{\beta}$ and $\bm{\nu}$. Then, the only remaining hyperparameter is the maximal basis degree $M_{\rm max}$. The pseudo-code of this new strategy is shown in Algorithm~\ref{alg:mop2}, where $\bm{\theta} \!\coloneqq\! {\bm{\nu}}^\frown\!\tilde{\beta}$ are the vectorized composite parameters concatenated from $\bm{\nu}$ and $\tilde{\beta}$. The initial value of $\bm{\nu}$ is set such that $\mathcal{C}_{N}\!\left(\bm{\nu}\right)$ match the exact nuclear geometry of the target system.

\begin{figure}[htb] % To make algorithm block float
\begin{algorithm}[H] % MUST use [H] to avoid compilation errors
    \caption{$\alpha$-bootstrap even-tempered basis optimization}\label{alg:mop2}
    \begin{algorithmic}[1]
        \renewcommand{\baselinestretch}{1.4}\selectfont
        \State Initialize $\alpha$, $\hat{\mathcal{E}}\!\left(\mathbb{M}\right)$, $M_{\rm max}$, $\bm{\nu}$
        \State $\tilde{\beta} \coloneqq 1/\alpha$
        \For {$M$ in $\left(1,\dots,M_{\rm max}\right)$}
            \If{$\tilde{\beta} < 1$}
                \State $\alpha \coloneqq \alpha\,\tilde{\beta}^M$
                \State $\tilde{\beta} \coloneqq 1/\tilde{\beta}$
            \EndIf
            \If{$\exists\,n\in\mathbb{Z}_{> 0}\ \text{such that}\ M=2^{\,n}$}
                \State $\alpha \coloneqq \alpha / \tilde{\beta}$
            \EndIf
            \State Initialize $\mathcal{G}^{\rm r}_{M}\!\left(\{\bm{\nu},\,\bm{\gamma}\}\right)$ with $\bm{\gamma}\coloneqq(\alpha,\,\tilde{\beta})$
            \State $E_{\rm HF} \coloneqq \hat{\mathcal{E}}\left[\mathcal{G}^{\rm r}_{M}\right]$, $i\coloneq0$, $\Delta\coloneqq\inf$
            \While {$i < i_{\rm max}$ and $\vrule\,\Delta\,\vrule > \delta$}
                \State ${
                    \setlength{\jot}{0.0em}
                    \begin{aligned}
                        \text{14.1)}\hspace{-2mm} & &i &\coloneqq\,i + 1 \\
                        \text{14.2)}\hspace{-2mm} & &{\bm{\nu}}^\frown\!\tilde{\beta} &\coloneqq\,{\rm{optimizer}}\!\left[\hat{\mathcal{E}},\,\partial \hat{\mathcal{E}}/\partial \bm{\theta}\!,\,\mathcal{G}^{\rm r}_{M}\right] \\
                        \text{14.3)}\hspace{-2mm} & &\bm{\gamma} &\coloneqq\, (\alpha,\,\tilde{\beta}) \\
                        \text{14.4)}\hspace{-2mm} & &\mathcal{G}^{\rm r}_{M} &\coloneqq\, \mathcal{G}^{\rm r}_{M}\!\left(\{\bm{\nu},\,\bm{\gamma}\}\right) \\ 
                        \text{14.5)}\hspace{-2mm} & &\Delta &\coloneqq\,E_{\rm HF} - \hat{\mathcal{E}}\left[\mathcal{G}^{\rm r}_{M}\right] \\
                        \text{14.6)}\hspace{-2mm} & &E_{\rm HF} &\coloneqq\,E_{\rm HF} - \Delta
                    \end{aligned}
                }$
            \EndWhile
            \State Store $\left(M,\,\bm{\nu},\,\bm{\gamma},\,E_{\rm HF}\right)$
        \EndFor
    \end{algorithmic}
\end{algorithm}
\end{figure}

\raggedbottom

\begin{figure}[h]
    \centering
    \includegraphics[width=\linewidth]{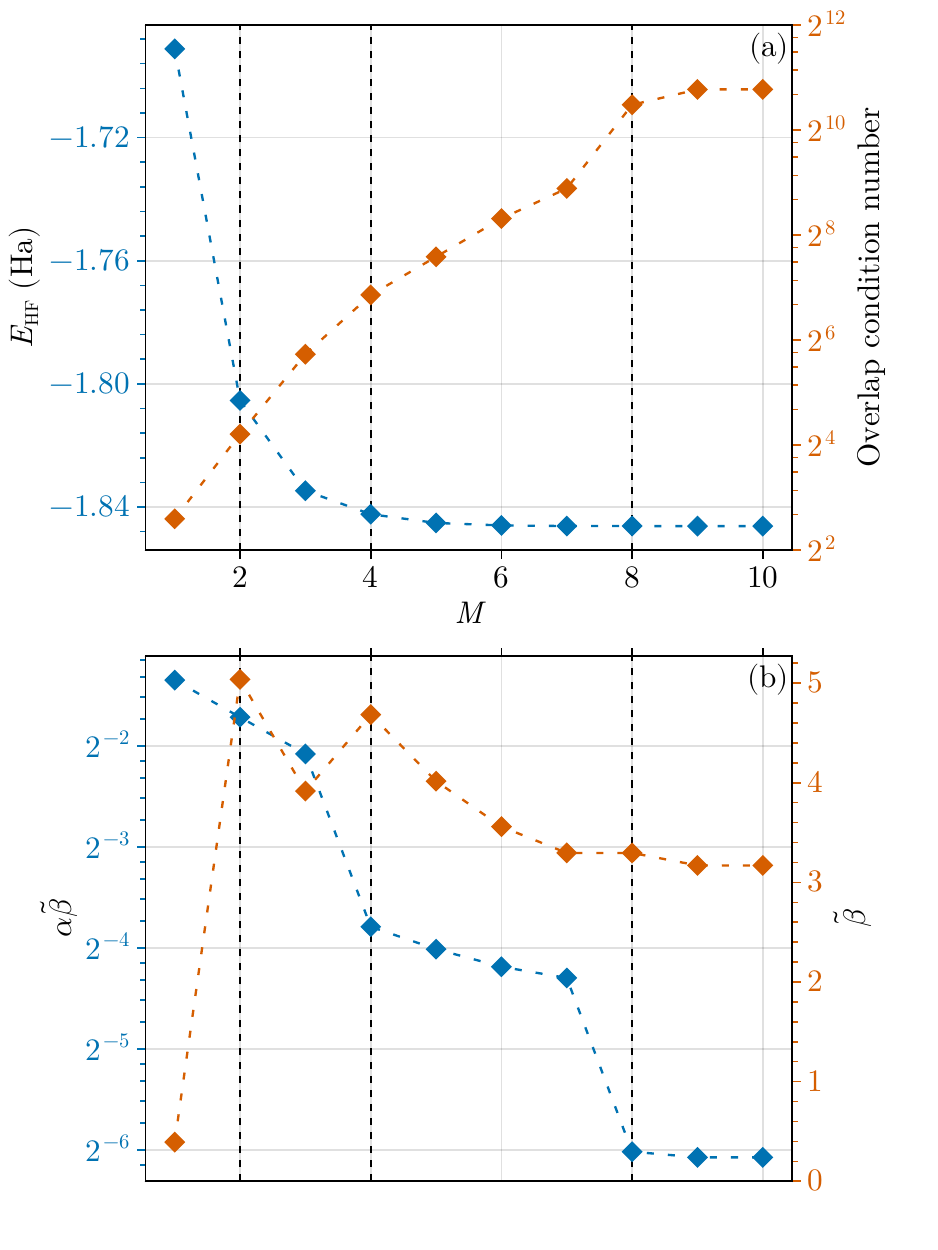}
    \vspace{-15pt}
    \caption{The convergence characteristics of a variationally optimized (through Algorithm \ref{alg:mop2}) reduced even-tempered basis set $\tilde{\mathcal{G}}^{\rm r}_M$  with respect to the basis degree $M$ in the case of H$_2$ at a bond length of $1.4$ a.u. (a) shows the restricted closed-shell Hartree--Fock energy (blue dashed line) for $\tilde{\mathcal{G}}^{\rm r}_M$ and the condition number of the corresponding overlap matrix (orange line). (b) shows the optimized values of the smallest basis-function exponent $\alpha\tilde{\beta}$ (blue dashed line) and the growth rate $\tilde{\beta}$ (orange dashed line). For (a) and (b), the vertical dashed lines cross the data points corresponding to the steps when $\alpha$ is updated.}
    \label{fig:H2_adapt}
\end{figure}

Aside from including the optimization of $\bm{\nu}$, the most significant addition in Algorithm \ref{alg:mop2} compared to Algorithm \ref{alg:mop1} is the bootstrap-based tuning procedure of $\alpha$ by lines 3--10. This procedure can be divided into two subroutines that are intermittently executed. The first subroutine (lines 3--7) effectively relabels the basis functions by restricting the domains of $\alpha$ and $\tilde{\beta}$ to be $(0, 1]$ and $(1, \infty)$, respectively. This subroutine preserves the single-particle function space that $\tilde{\mathcal{G}}^{\rm r}_M$ spans at the end of last optimization cycle, while resuming the optimization on a submanifold of $\hat{\mathcal{E}}\!\left(\mathbb{M}\right)$ to avoid degenerate local minima. Moreover, it helps compensate for the exponential growth of $\alpha$ by the polynomial growth of $\tilde{\beta}$. When the second subroutine (lines 8--10) is executed, $\alpha$ is divided by $\tilde{\beta}$, thus still bounded by $(0, 1]$. To avoid an exponential decay of $\alpha$ with respect to $M$, the update rate of $\alpha$ also follows an exponential decay. The two cases in which $\alpha$ is updated or remains the same as in the last cycle correspond to two directions for extending the basis set: appending a more diffuse basis function or a more localized one. In this way, the bootstrap initialization of the even-tempered basis set is more balanced compared to Algorithm \ref{alg:mop1}. To verify the numerical stability of Algorithm~\ref{alg:mop2}, we test it against the diatomic hydrogen at a typical bond length of $1.4$ a.u. Specifically, the exponent parameters $\bm{\gamma}$, the overlap condition number, and the HF energy for $\tilde{\mathcal{G}}^{\rm r}_M$ of different $M$ are plotted in FIG.~\ref{fig:H2_adapt}. 

First, FIG.~\ref{fig:H2_adapt}(a) shows that the HF energy with $\tilde{\mathcal{G}}^{\rm r}_M$ generated by Algorithm~\ref{alg:mop2} converges as $M$ increases. Particularly, each jump of the overlap condition number of the optimized basis set reflects the iteration cycle where $\alpha$ is updated at the beginning, which helps suppress the subsequent growth of the condition number. This update has a similar effect on the convergence of $\alpha\tilde{\beta}$ and $\tilde{\beta}$ in FIG.~\ref{fig:H2_adapt}(b). Specifically, when $\alpha$ is updated at the beginning of an optimization cycle, the resulting $\tilde{\beta}$ jumps (while $\alpha\tilde{\beta}$ drops), leading to a damped value decay in the subsequent cycles until the next update. Due to the consistent behavior across the performance indicators (energy and condition number) and the basis set parameters, one may choose one over the others as the criterion for obtaining the optimal minimum of $M$, depending on the required strictness for convergence.

\subsection{Dissociation of diatomic hydrogen}

\begin{figure}[htp]
    \centering
    \includegraphics[width=\linewidth]{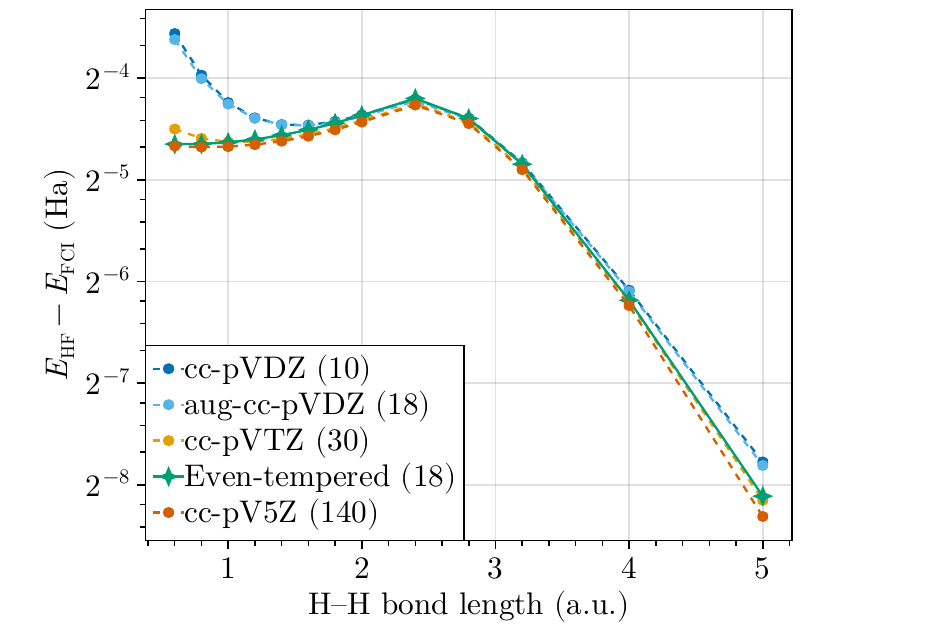}
    \vspace{-10pt}
    \caption{The unrestricted open-shell Hartree-Fock (UHF) energy errors of the H$_2$ molecule at different bond lengths with respect to the reduced variational even-tempered basis set ($\tilde{\mathcal{G}}^{\rm r}_{9}$) versus correlation-consistent basis sets. $\tilde{\mathcal{G}}^{\rm r}_{9}$ is generated by Algorithm~\ref{alg:mop2} also targeting the UHF functional. The size of each tested basis set is included in the parentheses after its name in the legend box.}
    \label{fig:H2_diss}
\end{figure}

We have demonstrated the performance scalability of Algorithm~\ref{alg:mop2} with respect to the basis degree $M$ for a fixed H$_2$ geometry. In this subsection, we test out its performance consistency at different H$_2$ bond lengths. Specifically, we compare $\tilde{\mathcal{G}}^{\rm r}_{9}$ generated by Algorithm~\ref{alg:mop2} against several correlation-consistent basis sets~\cite{dunning1989gaussian, kendall1992electron} (in the Cartesian representation) for producing the dissociation energy curve of H$_{2}$ on the HF level. The optimized parameters of the generated $\tilde{\mathcal{G}}^{\rm r}_{9}$ are listed in TABLE~\ref{tab:uhf-h2} in Appx.~\ref{app:1}, and results are shown in FIG.~\ref{fig:H2_diss}. cc-pV5Z (with optimized general contraction~\cite{pritchard2019new}) is also applied to the full configuration-interaction (FCI) computation (realized by PySCF~\cite{sun2020recent}) to estimate the ground-truth values near the CBS limit. The justification for choosing such a basis set is backed by numerical observations. For tested bond lengths, the energy differences between FCI/cc-pV5Z and FCI/aug-cc-pV5Z~\cite{dunning1989gaussian, kendall1992electron} are always below $\sim\!5\!\times\!10^{-4}$ Ha, approximately one third of the energy resolution at chemical accuracy.

Overall, the correlation energies for different basis sets exhibit the same monotonic trend in the medium-to-long bond region $[2,\,5]$ a.u. However, in $(0,\,1.8]$ a.u., only the correlation energies of $\tilde{\mathcal{G}}^{\rm r}_{9}$ and the near-complete cc-pV5Z still decrease, while the remaining tested basis sets produce qualitatively incorrect curves that increase the correlation energies. As a result, the HF energy of $\tilde{\mathcal{G}}^{\rm r}_{9}$ is lower than that of aug-cc-pVDZ and cc-pVTZ in this lower bond-length region. Notably, $\tilde{\mathcal{G}}^{\rm r}_{9}$ is of the same size as aug-cc-pVDZ (2$\times$9) and only contains S-subshell orbitals.

\begin{figure*}[htp]
    \centering
    \includegraphics[width=0.95\linewidth]{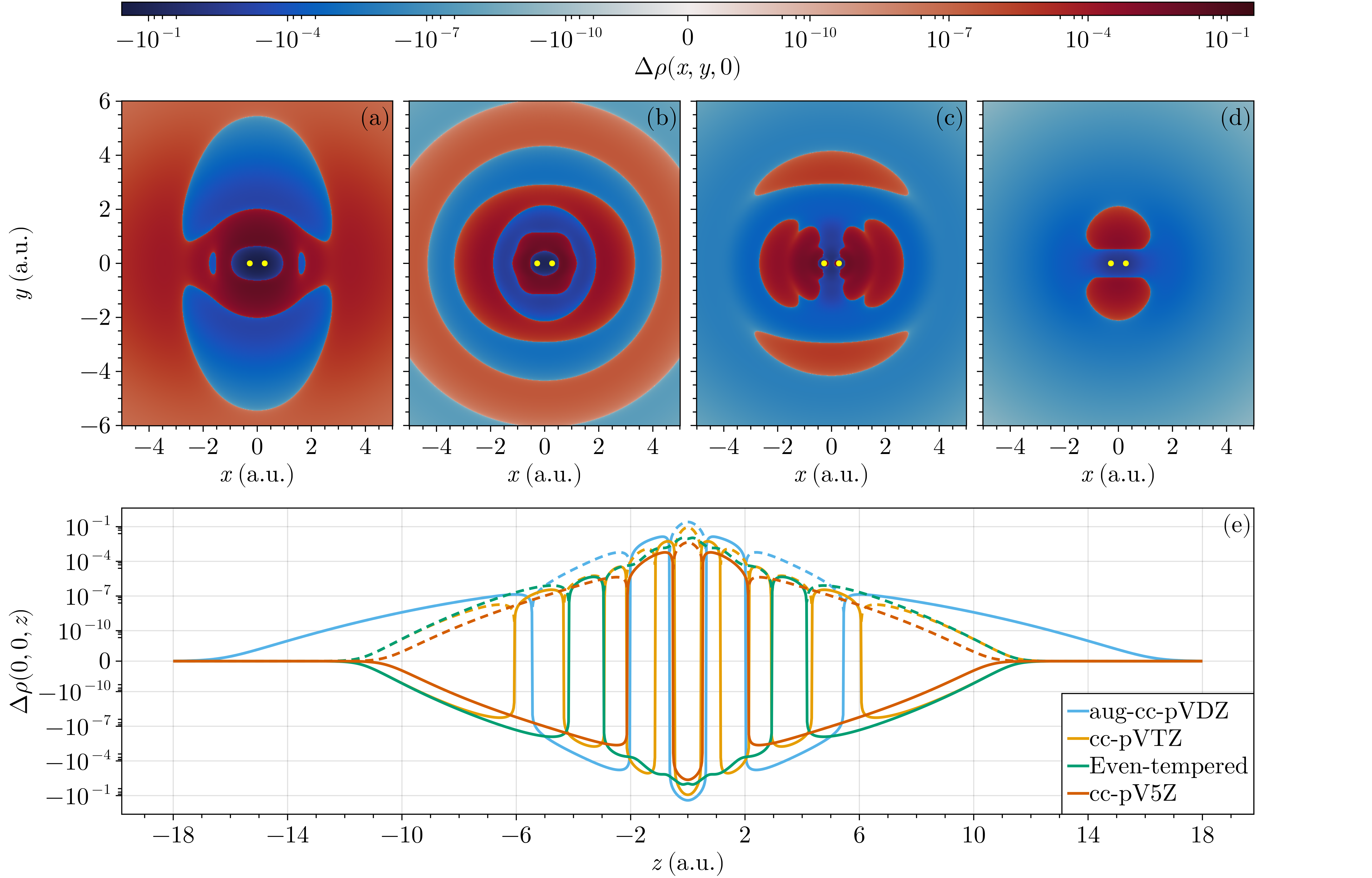}
    \vspace{-5pt}
    \caption{The differences between the charge densities on the HF level in various basis sets and the ground-truth charge density (FCI/cc-pV5Z) for the H$_2$ molecule at the bond length of $0.6$ a.u. Sub-figures (a--d) correspond to the density differences on the Cartesian $x$--$y$ plane ($z = 0$) of using aug-cc-pVDZ, cc-pVTZ, the even-tempered basis ($\tilde{\mathcal{G}}^{\rm r}_{9}$), and cc-pV5Z, respectively. The red areas represent overestimated regions, the blue areas represent underestimated regions, and the two yellow dots represent the nuclear positions. Sub-figure (e) shows the density differences along the $z$ axis ($x=y=0$) for the compared basis sets, where the solid lines represent the signed differences and the dashed lines represent the absolute differences.}
    \label{fig:H2_den}
\end{figure*}

To better estimate the discretization error of $\tilde{\mathcal{G}}^{\rm r}_{9}$ in the low bond-length region, we compare its Hartree--Fock charge density (HF density) against that of aug-cc-pVDZ and cc-pVTZ at the bond length of $0.6$ a.u., where the maximal performance divergence appears in FIG.~\ref{fig:H2_diss}. By taking the difference between the HF density of each compared basis set and the FCI/cc-pV5Z charge density (FCI density), we measure the density errors and visualize them in FIG.~\ref{fig:H2_den}. The HF density errors of cc-pV5Z are also included as a baseline reference for the HF approximation. It is possible to further integrate over the density difference and use the scalar result as a metric for the difference~\cite{carbo1980similar}. However, the distinct shapes of those density differences provide more information about the distribution of density errors, revealing inhomogeneous biases imposed by the incomplete basis sets. For instance, as shown by FIG.~\ref{fig:H2_den}(a--d), the density difference for each basis set forms multiple filled contours of distinct shapes in the $x$-$y$ plane. This geometric discrepancy indicates that the specific construction of incomplete basis sets imposes a non-uniformly distributed, yet spatially structured, overestimation (in red) or underestimation (in blue) in approximating the HF density at the CBS limit.

Focusing on the density errors along the $z$ axis, we also observe wavelet-like curves in FIG.~\ref{fig:H2_den}(e), which all have their respective main lobes centered at the origin, corresponding to the maximal amplitude of the single-point error. Restricting to the magnitude of the density errors (indicated by the flipped dashed lines), we can analyze the magnitude of the density error for each basis set more quantitatively. Compared to aug-cc-pVDZ, the error magnitude of $\tilde{\mathcal{G}}^{\rm r}_{9}$ decays more consistently and is especially lower near the origin and in the far region $(-\infty,\,-6)\cup(6,\,\infty)$. Compared to cc-pVTZ, $\tilde{\mathcal{G}}^{\rm r}_{9}$ becomes less advantageous. It produces produces lower errors in the region $(-4,\;4)$ a.u., higher errors in the region $(-6.2,\,-4) \cup (4,\;6.2)$ a.u., and asymptotically match the error of cc-pVTZ in the region $(-\infty,\,-6.2) \cup (6.2,\;\infty)$.

\subsection{Ground-state approximation of tetra-atomic hydrogen}\label{sec:H4}

\begin{figure}[htbp]
    \centering
    \includegraphics[width=\linewidth]{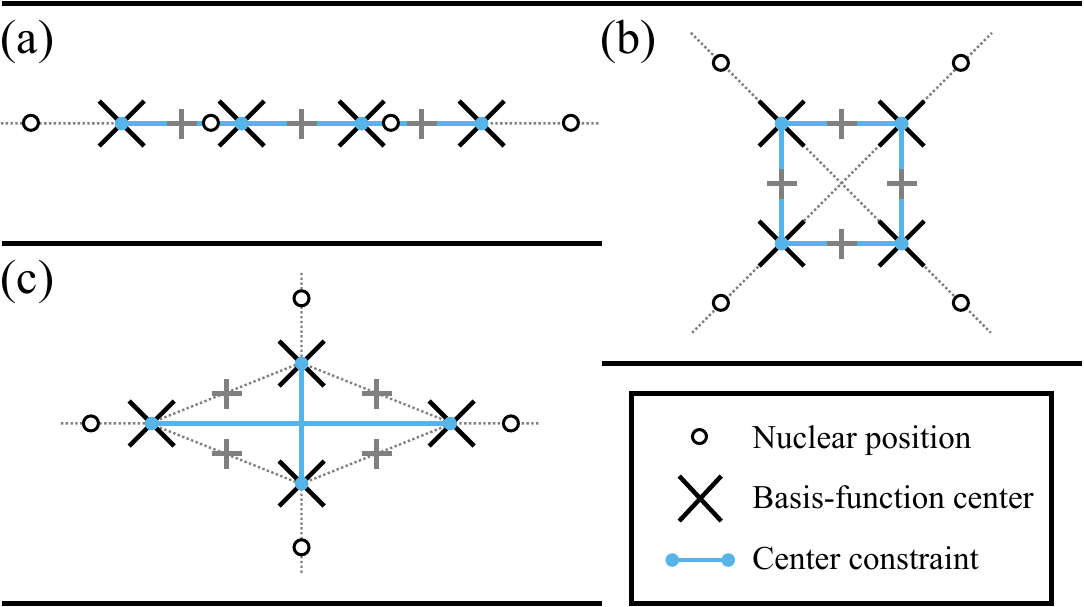}
    \caption{Three types of molecular H$_4$ used to test the variational even-tempered basis sets $\tilde{\mathcal{G}}^{\rm r}_M$ generated by Algorithm~\ref {alg:mop2}. For each system type, the respective correlated basis center $\mathcal{C}_{N}$, characterized by the ``center constraints'' (light blue edges), controls the basis-function centers of applied $\tilde{\mathcal{G}}^{\rm r}_M$. Sub-figure (a) represents the linear chain type, where the correlated basis center is constrained such that it stays in the same axis (marked as the dotted gray line) as the evenly spaced nuclear chain, shares a coincident midpoint, and has a tunable equal spacing parameter that determines the distance between adjacent basis-function centers. Sub-figure (b) represents the square planar type, where the correlated basis center is constrained such that it shares the same diagonal axes of symmetry (marked as the dotted gray lines) with the square formed by the nuclei, but has a tunable edge length parameter. Sub-figure (c) represents the rhombus type, where the correlated basis center is constrained such that the vertical (horizontal) center constraint stays in the same axis as the two vertically (horizontally) aligned nuclei and shares a midpoint, but has a separate tunable length parameter. The augmented centers used for the nested variational even-tempered basis sets are also marked by ``+'' in gray.}
    \label{fig:H4_layout}
\end{figure}

In the last subsection, we have demonstrated that, through Algorithm~\ref{alg:mop2}, we can generate variational even-tempered basis sets consisting solely of multi-center S-subshell orbitals that exhibit consistent performance over a range of bond lengths for the diatomic hydrogen molecule. In this subsection, we further test out the potential of Algorithm~\ref{alg:mop2} on molecular systems with relatively more complex geometries. Specifically, we compute the restricted closed-shell Hartree--Fock (RHF) energies with variational even-tempered basis set configurations against three types of tetra-atomic hydrogen (H$_4$) molecules: linear chain, square planar, and rhombus. The geometry configurations of these target systems are shown in FIG.~\ref{fig:H4_layout}.

We first directly applied Algorithm~\ref{alg:mop2} to generate variational even-tempered basis sets $\tilde{\mathcal{G}}^{\rm r}_M$ of three basis degrees with the correlated basis centers specified in FIG.~\ref{fig:H4_layout}. Their respective optimized parameters are listed in TABLE~\ref{tab:rhf_H4_simple} in Appx.~\ref{app:1}, and the corresponding RHF energies are shown in TABLE~\ref{tab:H4energy} (marked as ``Direct variational even-temepred''). Compared to aug-cc-pVDZ, which has the same number of basis functions, $\tilde{\mathcal{G}}^{\rm r}_{9}$ only produces lower RHF energy for the linear chain H$_4$ at bond lengths of $1.2$ a.u. and $1.6$ a.u. To verify if this underwhelming performance is caused by the lack of GTOs of higher angular momenta in $\tilde{\mathcal{G}}^{\rm r}_{9}$, we computed the RHF using subsets of aug-cc-pVDZ and cc-pVTZ that contain only S subshell orbitals (and S plus P subshell orbitals). Compared to the results of these subsets, which are also included in TABLE~\ref{tab:H4energy}, the directly generated variational even-tempered basis set does produce lower RHF energy when utilizing the same number of primitive S-subshell GTOs (e.g., $\tilde{\mathcal{G}}^{\rm r}_{6}$ vs. the S-subshell part of aug-cc-pVDZ). However, it does not necessarily offer any advantage in terms of basis-set size. This is because in aug-cc-pVDZ and cc-pVTZ, the primitive GTOs can be combined to form contracted GTOs. Moreover, the inclusion of P-subshell GTOs indeed significantly lowers the energy, which cannot be achieved with such a direct even-tempered construction. 

\begin{table*}[htbp]
    \setlength{\tabcolsep}{2pt}
    \renewcommand{\arraystretch}{1.6}
    \centering
    \begin{tabular}{cccccccccc}
        \toprule[1.5pt]
        \multicolumn{3}{c}{\multirow{2}{*}{Basis set}} & \multicolumn{6}{c}{H$_4$ RHF energy excluding nuclear repulsions (Ha)} \\ \cline{4-9} 
        \multicolumn{3}{c}{} & \multicolumn{3}{c}{Linear chain} & \multicolumn{2}{c}{Square planar} & \multicolumn{1}{c}{Rhombus} \\ \hline
        \multicolumn{2}{c}{Type} & \multicolumn{1}{c}{Basis degree} & \multicolumn{1}{c}{1.2 a.u.} & \multicolumn{1}{c}{1.6 a.u.} & \multicolumn{1}{c}{2.0 a.u.} & \multicolumn{1}{c}{2.0 a.u.} & \multicolumn{1}{c}{2.4 a.u.} & \multicolumn{1}{c}{2.2 a.u.} \\ \midrule[1.0pt]
        % \multicolumn{2}{c}{STO-6G}                & 6$\,\to\,$1  & $-$5.62318 & $-$4.85170 & $-$4.25536 & $-$4.49982 & $-$4.04413 & $-$4.40613 \\ \hline
        \multicolumn{2}{c}{\multirow{3}{*}{\begin{tabular}[c]{@{}c@{}}Direct variational\\[-5pt]even-tempered\end{tabular}}} 
                                                  & 3 & $-$5.61693 & $-$4.86449 & $-$4.30367 & $-$4.63276 & $-$4.17920 & $-$4.54378 \\
        \multicolumn{2}{c}{}                      & 6 & $-$5.67733 & $-$4.88849 & $-$4.31855 & $-$4.65320 & $-$4.20025 & $-$4.56171 \\
        \multicolumn{2}{c}{}                      & 9 & $-$5.67814 & $-$4.88905 & $-$4.31922 & $-$4.65369 & $-$4.20099 & $-$4.56200 \\ \hline
        \multicolumn{2}{c}{\multirow{3}{*}{\begin{tabular}[c]{@{}c@{}}Nested variational\\[-5pt]even-tempered\end{tabular}}} 
                                                  & (6, 1) & $-$5.67905 & $-$4.89083 & $-$4.32096 & $-$4.65734 & $-$4.20421 & $-$4.56993 \\
        \multicolumn{2}{c}{}                      & (6, 2) & $-$5.67918 & $-$4.89126 & $-$4.32138 & $-$4.65816 & $-$4.20474 & $-$4.57030 \\
        \multicolumn{2}{c}{}                      & (6, 3) & $-$5.67931 & $-$4.89141 & $-$4.32231 & $-$4.65837 & $-$4.20538 & $-$4.57116 \\ \hline
        \multirow{2}{*}{aug-cc-pVDZ} & S subshell & 6$\,\to\,$3 & $-$5.65354 & $-$4.88051 & $-$4.31547 & $-$4.64378 & $-$4.19302 & $-$4.54578 \\
         & Total (S$+$P)                                  & 12$\,\to\,$9 & $-$5.66054 & $-$4.88549 & $-$4.32032 & $-$4.65981 & $-$4.20704 & $-$4.57663 \\ \hline
        \multirow{3}{*}{cc-pVTZ} & S subshell     & 7$\,\to\,$3 & $-$5.67545 & $-$4.88672 & $-$4.31724 & $-$4.64549 & $-$4.19539 & $-$4.54835 \\
         & S$+$P subshells                        & 13$\,\to\,$9 & $-$5.67935 & $-$4.89210 & $-$4.32297 & $-$4.66207 & $-$4.20831 & $-$4.57838 \\
         & Total (S$+$P$+$D)                                  & 19$\,\to\,$15 & $-$5.67945 & $-$4.89218 & $-$4.32331 & $-$4.66251 & $-$4.20859 & $-$4.57885 \\
         \bottomrule[1.5pt]
    \end{tabular}
    \caption{Comparison between even-tempered basis sets generated by Algorithm~\ref{alg:mop2} and several atomic (Cartesian) Gaussian basis sets in being applied to computing the RHF energy for various geometries of H$_4$ molecules. For the linear chain type, three interatomic distances (i.e., bond lengths) are tested: $1.2$ a.u., $1.6$ a.u., and $2.0$ a.u. For the square planar type, two edge lengths are tested: $2.0$ a.u. and $2.4$ a.u. Additionally, a rhombus type with a $60^{\circ}$ angle and an edge length of $2.2$ a.u. is tested. For each ``nested variational even-tempered basis set'' configuration, its basis degrees are specified by a pair of integers. The first represents the basis degree of its based ``direct variational even-tempered'' basis set, and the second represents the number of even-tempered basis functions added at each augmented center. For the atomic Gaussian basis sets (and respective subsets divided by angular momentum subshells), ``basis degree'' shows the number of orbitals at each nuclear position before and after (indicated by ``$\to$'') the contraction of primitive GTOs.}
    \label{tab:H4energy}
\end{table*}

To address the limitations of S-subshell orbitals centered on correlated basis centers within the current construction and optimization framework, we introduce the notion of \textbf{nested variational even-tempered basis sets}. They are generated based on the already-optimized $\tilde{\mathcal{G}}^{\rm r}_M$, which are re-categorized as the \textbf{direct variational even-tempered basis sets}. We extend $\tilde{\mathcal{G}}^{\rm r}_M$ by adding even-tempered basis functions parameterized by a new $\gamma$ on the augmented centers that are at midpoints between its basis-function centers, then optimize $\gamma$ by a procedure similar to Algorithm~\ref{alg:mop2}. We can characterize such a nested construction by a pair of basis degrees. The first basis degree inherits from the original $\tilde{\mathcal{G}}^{\rm r}_M$, and the second basis degree specifies the number of even-tempered basis functions located on each augmented center. For the previously tested H$_4$ geometries, we define the augmented centers in FIG.~\ref{fig:H4_layout} by gray markers shaped like ``+''. We used these augmented centers to generate nested variational even-tempered basis sets based on $\tilde{\mathcal{G}}^{\rm r}_{6}$. The optimized parameters of the appended basis subsets with different augmented basis degrees are listed in TABLE~\ref{tab:rhf_H4_nest} in Appx.~\ref{app:1}, and the RHF energies of the corresponding nested basis sets are shown in rows 3--6 of TABLE~\ref{tab:H4energy}. Even just adding even-tempered basis functions with basis degree $M\!=\!1$, the performance of the even-tempered basis set is already significantly improved. At a basis-degree pair of (6, 2), the nested variational even-tempered basis set contains fewer basis functions than aug-cc-pVDZ for all H$_4$ geometries. Yet, this configuration produces lower RHF energy for the linear chain H$_4$ and comparable energies (difference below $2.5\!\times\!10^{-3}$ Ha) for the square H$_4$. To confirm that for each tested H$_4$ geometry, all the HF states corresponding to the compared RHF energies have the same spatial symmetry, we also computed their respective charge densities. The visualized comparisons of these charge densities are included in Appx.~\ref{app:2}.

\section{Conclusions and Discussions}\label{sec:5}

In this paper, we revisit the even-tempered basis functions and use them as a direct approach to construct molecular orbitals for electronic ground-state wavefunctions. Based on the conventional even-tempered formalism, we propose system-oriented basis set designs with variational and numerically stable optimization procedures.

In Sec.~\ref{sec:2}, we first propose a reduced formalism of even-tempered basis sets and define the (reduced) variational even-tempered basis sets $\tilde{\mathcal{G}}^{\rm r}_M$ with tunable parameters optimized\cite{wang2023basis} against the electronic ground-state energy. Next, in Sec.~\ref{sec:3}, we apply it to the atomic hydrogen to quantify the correlation between the two parameters $\alpha$ and $\tilde{\beta}$ that determine the optimal exponent coefficients of the even-tempered basis functions. We show that even when these two parameters are not optimized simultaneously (through Algorithm~\ref{alg:mop1}), one can still systematically achieve the exponential decay of the ground-state energy error (see FIG.~\ref{fig:H_energy}). This numerical observation rationalizes the hierarchical separation of tuning $\alpha$ and $\tilde{\beta}$. We further note the numerical instability of conventionally optimized even-tempered basis sets, which arises from the exponential scaling of the overlap matrix's condition number (see Fig.~\ref{fig:H_cond}). We propose a simplified formula for generating $\tilde{\mathcal{G}}^{\rm r}_M$ in Equation~(\ref{eq:et_map}), where only $\tilde{\beta}$ is actively optimized for the hydrogen atom. Not only does the accuracy of this simplified version of $\tilde{\mathcal{G}}^{\rm r}_M$ match the conventional approach across different basis degree $M$, but it also presents better numerical scalability.

After improving the optimization of atomic even-tempered basis sets, we consider the direct generation of multi-center even-tempered basis sets for molecular systems. In Sec.~\ref{sec:4}, we propose a two-level optimization strategy (in Algorithm~\ref{alg:mop2}) that optimizes both the exponent coefficients and the basis-function centers. In particular, the basis-function centers are specified by the ``correlated basis centers'' defined in Equation~(\ref{eq:nc}), which are correlated through primitive parameters $\bm{\nu}$. By such construction, these centers remain floating (non-fixed) while preserving a desired geometry when $\bm{\nu}$ is being optimized. To test the performance of Algorithm~\ref{alg:mop2}, we first apply it to the diatomic hydrogen (H$_2$) molecule and compare its dissociation curve with those of correlation-consistent basis sets. We observe a consistent error across the region where the bond length exceeds $1.8$ a.u. As for the region where the bond length is below $1.8$ a.u., the variational even-tempered basis set of degree nine ($\tilde{\mathcal{G}}^{\rm r}_{9}$) produces lower errors than aug-cc-pVDZ (which has the same basis-set size) and cc-pVTZ. To verify the discretization advantage of $\tilde{\mathcal{G}}^{\rm r}_{9}$ over aug-cc-pVDZ on constructing proper molecular orbitals, we further compare their corresponding electron charge densities at the bond length of $0.6$ a.u., and the results are shown in FIG.~\ref{fig:H2_den}. Particularly, we show that $\tilde{\mathcal{G}}^{\rm r}_{9}$ more efficiently describes the charge density in the compressed and stretched bond regions than the traditional atomic Gaussian basis sets. Next, we test the performance of variational even-tempered basis sets generated by Algorithm~\ref{alg:mop2} on various tetra-atomic hydrogen molecules (H$_4$). Based on the test results in TABLE~\ref{tab:H4energy}, we conclude that for systems more complicated than homonuclear diatomic molecules, simply placing on even-tempered basis functions on centers that maintain the spatial symmetry of the nuclear positions is not enough to produce energy with the level of accuracy from atomic Gaussian basis sets. However, as we add more even-tempered basis functions on secondary basis-function centers (i.e., the augmented centers), the performance of the even-tempered basis set can be significantly improved through further partial variational optimization.

Now, we look back at the question in Sec.~\ref{sec:1}: How efficiently can a system-oriented even-tempered basis set encode the electronic ground-state information beyond atomic systems? We have demonstrated a few cases in which even-tempered basis sets can provide a more efficient representation of molecular orbitals at the mean-field level. However, based on the results for the H$_4$ systems, we believe there is room for further improvement. For instance, we can try including even-tempered basis functions of angular momentum higher than the S subshell. Moreover, we would like to compare the benefits of such additions to those of the nested approaches based on S-subshell GTOs at augmented centers. Specifically, we can implement more sophisticated correlated basis centers for the augmented S-subshell GTOs and compare their contribution against directly adding even-tempered GTOs in the higher angular momentum subshells. Testing these potential modifications across a wider range of molecules can help us better understand the limitations of the current framework and improve it.

On the one hand, our work provides a minimal starting point for developing more holistic electronic-structure methods without relying on tabulated empirical data. This system-oriented generation of basis sets on the Hartree-Fock level provides a more direct connection between basis-set discretizations and the formation of molecular orbitals for many-electron systems than direct linear combinations of atomic basis sets. On the other hand, more in-depth questions also need to be answered in order to assess the practicality of using system-oriented even-tempered basis sets for electronic structure computation. Is there a more systematic way to nest multiple even-tempered basis sets based on the geometry of the target system? How should one determine the basis degree for each subset of the even-tempered basis functions before the variational optimization procedure? What is the average optimization complexity of the parameter space of the even-tempered basis sets? Admittedly, the conventional atomic Gaussian basis sets share some similar concerns, such as determining the cardinal number ($n$ in cc-pV$n$Z) of the correlation-consistent basis sets. However, compared to the system-oriented approach, they have little to no overhead of reconfiguring the construction of the final basis set. Therefore, for data-free but not parameter-free basis sets like variational even-tempered basis sets, the main challenge for their applicability within the computational chemistry community does not only depend on whether they can efficiently encode the electronic ground-state information, but also depends on how easily and consistently they can do it. In addition to further improving the upper-bound performance of system-oriented even-tempered basis sets, we hope that future work will also explore the practicality of incorporating this basis-set design framework into a more realistic working pipeline for electronic structure computation.

\begin{acknowledgments}
W.W. thanks Brent Harrison for the helpful suggestions on visualizing the density differences in FIG.\ref{fig:H2_den}. W.W. and J.D.W. acknowledge support by the U.S. Department of Energy, Office of Basic Energy Sciences, under Award DE-SC0019374. W.W. and J.D.W. were also partially supported by the ARO grant W911NF2410043. J.D.W. holds concurrent appointments at Dartmouth College and as an Amazon Visiting Academic. This paper describes work performed at Dartmouth College and is not associated with Amazon.
\end{acknowledgments}

\bibliography{main}

\appendix

\clearpage
\onecolumngrid

\section{Optimized basis set parameters}\label{app:1}

\begin{table*}[htp]
  \centering
  \setlength{\tabcolsep}{2pt}
  \renewcommand{\arraystretch}{1.1}
  \vspace{-10pt}

  \begin{minipage}[t]{0.4\textwidth}
  \vspace{0pt}
  \centering
  \begin{tabular*}{\linewidth}{@{\extracolsep{\fill}} c c c c @{} }
    \toprule[1.5pt]
    $\alpha$ & $\tilde{\beta}$ & $M$ & Energy (Ha)\\
    \midrule[1.0pt]
    \multirow{3}{*}{1} & 0.393140 & 2 & $-$0.44916 \\
    & 0.667944 & 4 & $-$0.47852 \\
    & 0.797569 & 6 & $-$0.48864 \\
    \midrule
    \multirow{4}{*}{2} & 0.274151 & 2 & $-$0.46806 \\
    & 0.519582 & 4 & $-$0.48597 \\
    & 0.662166 & 6 & $-$0.49235 \\
    & 0.748984 & 8 & $-$0.49524 \\
    \midrule
    \multirow{6}{*}{4} & 0.204947 & 2 & $-$0.48093 \\
    & 0.417854 & 4 & $-$0.49206 \\
    & 0.562148 & 6 & $-$0.49560 \\
    & 0.657208 & 8 & $-$0.49722 \\
    & 0.722829 & 10 & $-$0.49809 \\
    \midrule
    \multirow{7}{*}{8}  & 0.157000 & 2 & $-$0.48574 \\
    & 0.345140 & 4 & $-$0.49596 \\
    & 0.484849 & 6 & $-$0.49773 \\
    & 0.583781 & 8 & $-$0.49855 \\
    & 0.655322 & 10 & $-$0.49900 \\
    & 0.708195 & 12 & $-$0.49926 \\
    \midrule
    \multirow{7}{*}{16} & 0.119268 & 2 & $-$0.48345 \\
    & 0.292807 & 4 & $-$0.49800 \\
    & 0.424110 & 6 & $-$0.49892 \\
    & 0.523709 & 8 & $-$0.49930 \\
    & 0.597673 & 10 & $-$0.49951 \\
    & 0.654368 & 12 & $-$0.49964 \\
    & 0.698894 & 14 & $-$0.49972 \\
    
    \bottomrule[1.5pt]
    \end{tabular*}
\end{minipage}
    \hfill
    \begin{minipage}[t]{0.4 \textwidth}
    \vspace{0pt}
    \centering
    \begin{tabular*}{\linewidth}{@{\extracolsep{\fill}} c c c c @{} }
      \toprule[1.5pt]
      $\alpha$ & $\tilde{\beta}$ & $M$ & Energy (Ha)\\
      \midrule[1.0pt]
    \multirow{9}{*}{32} & 0.088965 & 2 & $-$0.47608 \\
    & 0.253055 & 4 & $-$0.49873 \\
    & 0.375734 & 6 & $-$0.49951 \\
    & 0.473296 & 8 & $-$0.49968 \\
    & 0.548382 & 10 & $-$0.49978 \\
    & 0.607553 & 12 & $-$0.49983 \\
    & 0.655120 & 14 & $-$0.49987 \\
    & 0.693002 & 16 & $-$0.49990 \\
    \midrule
    \multirow{10}{*}{64} & 0.065135 & 2 & $-$0.46606 \\
    & 0.219475 & 4 & $-$0.49847 \\
    & 0.337393 & 6 & $-$0.49977 \\
    & 0.430413 & 8 & $-$0.49986 \\
    & 0.506717 & 10 & $-$0.49990 \\
    & 0.565911 & 12 & $-$0.49993 \\
    & 0.616830 & 14 & $-$0.49994 \\
    & 0.655119 & 16 & $-$0.49995 \\
    & 0.686586 & 18 & $-$0.49996 \\
    \midrule
    \multirow{12}{*}{128} & 0.046955 & 2 & $-$0.45571 \\
    & 0.189911 & 4 & $-$0.49732 \\
    & 0.305748 & 6 & $-$0.49984 \\
    & 0.394849 & 8 & $-$0.49994 \\
    & 0.467925 & 10 & $-$0.49996 \\
    & 0.527660 & 12 & $-$0.49997 \\
    & 0.579845 & 14 & $-$0.49998 \\
    & 0.616841 & 16 & $-$0.49998 \\
    & 0.648062 & 18 & $-$0.49998 \\
    & 0.672647 & 20 & $-$0.49999 \\
    \bottomrule[1.5pt]
  \end{tabular*}
  \end{minipage}

  \vspace{-5pt}

  \caption{The values of $\tilde{\beta}$ output by Algorithm~\ref{alg:mop1} under a given $\alpha$ to generate the reduced variational even-tempered basis sets $\tilde{\mathcal{G}}^{\rm r}_M$ for the hydrogen atom. The rightmost column lists the respective ground-state energies.}
  \label{tab:rhf-hydrogen}
  \vspace{-10pt}
\end{table*}

\begin{table*}[htp]
  \centering
  \setlength{\tabcolsep}{2pt}
  \renewcommand{\arraystretch}{1.1}
  \begin{tabular*}{0.8 \linewidth}{@{\extracolsep{\fill}} c c c c c @{} }
    \toprule[1.5pt]
    H$_2$ bond length (a.u.) & $\alpha$ & $\tilde{\beta}$ & $\bm{\nu}$ (a.u.) &  Energy (Ha)\\
    \midrule[1.0pt]
    0.6 & 0.054307 & 2.644041 & 0.552297 & $-$2.39608 \\
    0.8 & 0.026985 & 2.788918 & 0.737008 & $-$2.22986 \\
    1.0 & 0.013222 & 2.865272 & 0.924141 & $-$2.08395 \\
    1.2 & 0.007695 & 3.022383 & 1.114314 & $-$1.95690 \\
    1.4 & 0.004678 & 3.170136 & 1.307021 & $-$1.84620 \\
    1.6 & 0.003054 & 3.081738 & 1.501414 & $-$1.74944 \\
    1.8 & 0.002135 & 3.113571 & 1.698202 & $-$1.66445 \\
    2.0 & 0.001594 & 3.206404 & 1.897016 & $-$1.58941 \\
    2.4 & 0.000955 & 3.936617 & 2.299774 & $-$1.46442 \\
    2.8 & 0.000745 & 4.052290 & 2.761710 & $-$1.38011 \\
    3.2 & 0.000581 & 4.206682 & 3.181930 & $-$1.32327 \\
    4.0 & 0.000461 & 3.458980 & 3.994340 & $-$1.25240 \\
    5.0 & 0.000303 & 3.675462 & 4.998815 & $-$1.20001 \\
    \bottomrule[1.5pt]
    \end{tabular*}
    \vspace{-5pt}
     \caption{The values of $\alpha$, $\tilde{\beta}$, and the correlated-cluster parameters ($\bm{\nu}$) output by Algorithm~\ref{alg:mop2} to generate the reduced variational even-tempered basis sets $\tilde{\mathcal{G}}^{\rm r}_{9}$ for the molecular H$_2$ at different bond lengths. The rightmost column lists the respective UHF energies (excluding nuclear repulsion), as the UHF functional was used as the objective in Algorithm~\ref{alg:mop2}.}
  \label{tab:uhf-h2}
  \vspace{-15pt}
\end{table*}

\begin{table*}[htp]
  \centering
  \setlength{\tabcolsep}{2pt}
  \renewcommand{\arraystretch}{1.2}
  \begin{tabular*}{0.95\linewidth}{@{\extracolsep{\fill}} c c c c c c c @{} }
    \toprule[1.5pt]
    Type & Geometry & $\alpha$ & $\tilde{\beta}$ & $\bm{\nu}$ (a.u.) & $M$ & Energy (Ha)\\
    \midrule[1.0pt]
    \multirow{9}{*}{Linear chain} & \multirow{3}{*}{1.2 a.u.} & 0.117587 & 2.715992 & 1.161747 & 3 & $-$5.61693 \\
    && 0.043294 & 3.154529 & 1.179023 & 6 & $-$5.67733\\
    && 0.014507 & 3.010633 & 1.180780 & 9 & $-$5.67814 \\
    \cline{2-7} 
    & \multirow{3}{*}{1.6 a.u.} & 0.059646 & 3.991194 & 1.563404 & 3 & $-$4.86449 \\
    && 0.014944 & 3.630591 & 1.576631 & 6 & $-$4.88849 \\
    && 0.004282 & 3.478584 & 1.578644 & 9 & $-$4.88905 \\ 
    \cline{2-7} 
    & \multirow{3}{*}{2.0 a.u.} & 0.035272 & 4.865904 & 1.961840 & 3 & $-$4.30367 \\
    && 0.007249 & 3.903687 & 1.976556 & 6 & $-$4.31855 \\
    && 0.001929 & 3.771001 & 1.978518 & 9 & $-$4.31922 \\
    \midrule 
    \multirow{6}{*}{Square planar} & \multirow{3}{*}{2.0 a.u.} & 0.032225 & 4.632960 & 1.890077 & 3 & $-$4.63276 \\
    && 0.006956 & 3.796694 & 1.924379 & 6 & $-$4.65320 \\
    && 0.001869 & 3.732178 & 1.927855 & 9 & $-$4.65369 \\
    \cline{2-7}
    & \multirow{3}{*}{2.4 a.u.} & 0.022950 & 4.988391 & 2.292533 & 3 & $-$4.17920 \\ 
    && 0.004601 & 3.949170 & 2.335201 & 6 & $-$4.20025 \\
    && 0.001194 & 3.860293 & 2.339247 & 9 & $-$4.20099 \\
    \midrule
    \multirow{3}{*}{Rhombus} & \multirow{3}{*}{(60$^{\circ}$, 2.2 a.u.)} & 0.028346 & 4.774232 & (3.589486, 1.996967) & 3 & $-$4.54378\\
    && 0.005937 & 3.815942 & (3.655913, 2.045451) & 6 & $-$4.56171 \\
    && 0.001583 & 3.771706 & (3.661470, 2.047756) & 9 & $-$4.56200 \\
    \bottomrule[1.5pt]
    \end{tabular*}
    \vspace{-5pt}
     \caption{The values of $\alpha$, $\tilde{\beta}$, and the correlated-cluster parameters ($\bm{\nu}$) output by Algorithm~\ref{alg:mop2} to generate the direct variational even-tempered basis sets $\tilde{\mathcal{G}}^{\rm r}_M$ for the molecular H$_4$ geometries listed in Table \ref{tab:H4energy}. The rightmost column lists the respective RHF energies (excluding nuclear repulsions), as the RHF functional was used as the objective in Algorithm~\ref{alg:mop2}.}
  \label{tab:rhf_H4_simple}
\end{table*}

\begin{table*}[htp]
  % \vspace{-5pt}
  \centering
  \setlength{\tabcolsep}{2pt}
  \renewcommand{\arraystretch}{1.2}
  \begin{tabular*}{0.85\linewidth}{@{\extracolsep{\fill}} c c c c c c @{} }
    \toprule[1.5pt]
    Type & Geometry & $\alpha$ & $\tilde{\beta}$ & $M$ & Energy (Ha)\\
    \midrule[1.0pt]
    \multirow{9}{*}{Linear chain}  & \multirow{3}{*}{1.2 a.u.}  & 1 & 1.522179 & 1 & $-$5.67905 \\
                                             && 0.105112 & 4.235839 & 2 & $-$5.67918\\
                                             && 0.105112 & 2.866763 & 3 & $-$5.67931\\
    \cline{2-6}
                                              & \multirow{3}{*}{1.6 a.u.}  & 1 & 0.695538 & 1 & $-$4.89083 \\
                                             && 0.230038 & 2.377820 & 2 & $-$4.89126 \\
                                             && 0.230038 & 1.839277 & 3 & $-$4.89141 \\
    \cline{2-6}
                                              & \multirow{3}{*}{2.0 a.u.}  & 1 & 0.741479 & 1 & $-$4.32096 \\
                                             && 0.215785 & 1.768776 & 2 & $-$4.32138 \\
                                             && 0.215785 & 1.370233 & 3 & $-$4.32231 \\
    \midrule 
    \multirow{6}{*}{Square planar} & \multirow{3}{*}{2.0 a.u.} & 1 & 0.372993 & 1 & $-$4.65734 \\
                                             && 0.107241 & 2.396058 & 2 & $-$4.65816 \\
                                             && 0.107241 & 1.764328 & 3 & $-$4.65837 \\
    \cline{2-6}
                                              & \multirow{3}{*}{2.4 a.u.} & 1 & 0.280993 & 1 & $-$4.20421 \\ 
                                             && 0.138664 & 1.639747 & 2 & $-$4.20474 \\
                                             && 0.138664 & 1.332234 & 3 & $-$4.20538 \\
    \midrule
    \multirow{3}{*}{Rhombus}       & \multirow{3}{*}{(60$^{\circ}$, 2.2 a.u.)}       & 1 & 0.312303 & 1 & $-$4.56993\\
                                             && 0.190375 & 1.469011 & 2 & $-$4.57030 \\
                                             && 0.190375 & 1.213282 & 3 & $-$4.57116 \\
    \bottomrule[1.5pt]
    \end{tabular*}
    \vspace{-5pt}
     \caption{The values of $\alpha$ and $\tilde{\beta}$ output by Algorithm~\ref{alg:mop2} to generate the additional even-tempered basis functions located on the augmented centers (shown in FIG.~\ref{fig:H4_layout}) to construct the nested variational even-tempered basis sets based on  $\tilde{\mathcal{G}}^{\rm r}_{6}$ for the molecular H$_4$ geometries listed in Table \ref{tab:H4energy}. Particularly, for basis degree $M\!=\!1$, the value of $\alpha$ is always normalized to one so that the corresponding $\tilde{\beta}$ directly represents the exponent coefficient of the single S-subshell GTO at each augmented center. The rightmost column lists the respective RHF energies (excluding nuclear repulsions), as the RHF functional was used as the objective in Algorithm~\ref{alg:mop2}.}
  \label{tab:rhf_H4_nest}
  \vspace{-10pt}
\end{table*}

\newpage

\section{Charge densities of tetra-atomic hydrogen}\label{app:2}

\begin{figure}[thb]
  \centering
  \begin{minipage}[t]{0.48\linewidth}
    \centering
    \includegraphics[width=\linewidth]{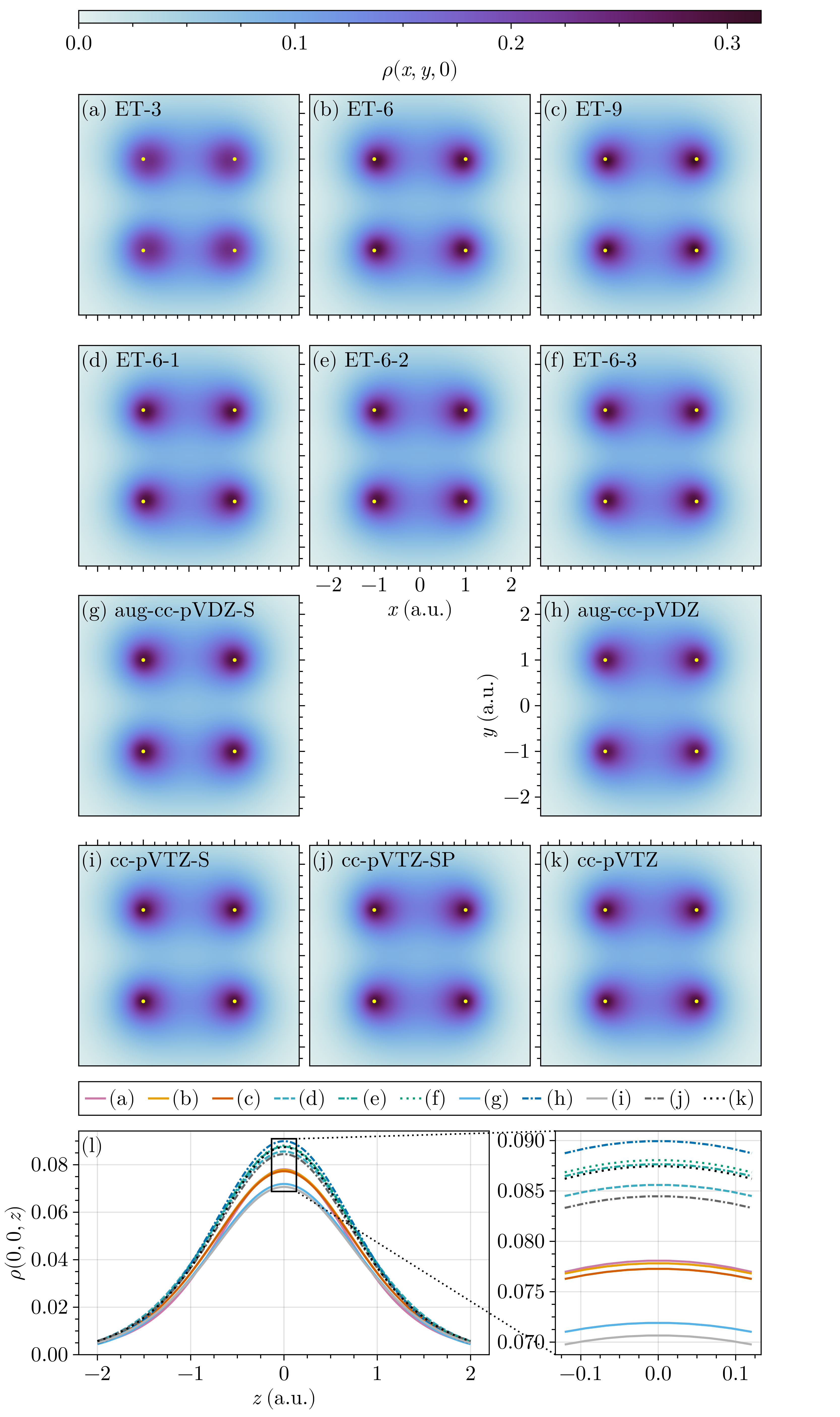}
    \vspace{-15pt}
    \captionof{figure}{Square planar ($2.0$ a.u.).}
    \label{fig:den_H4sqrt2p0}
  \end{minipage}\hfill
  \begin{minipage}[t]{0.48\linewidth}
    \centering
    \includegraphics[width=\linewidth]{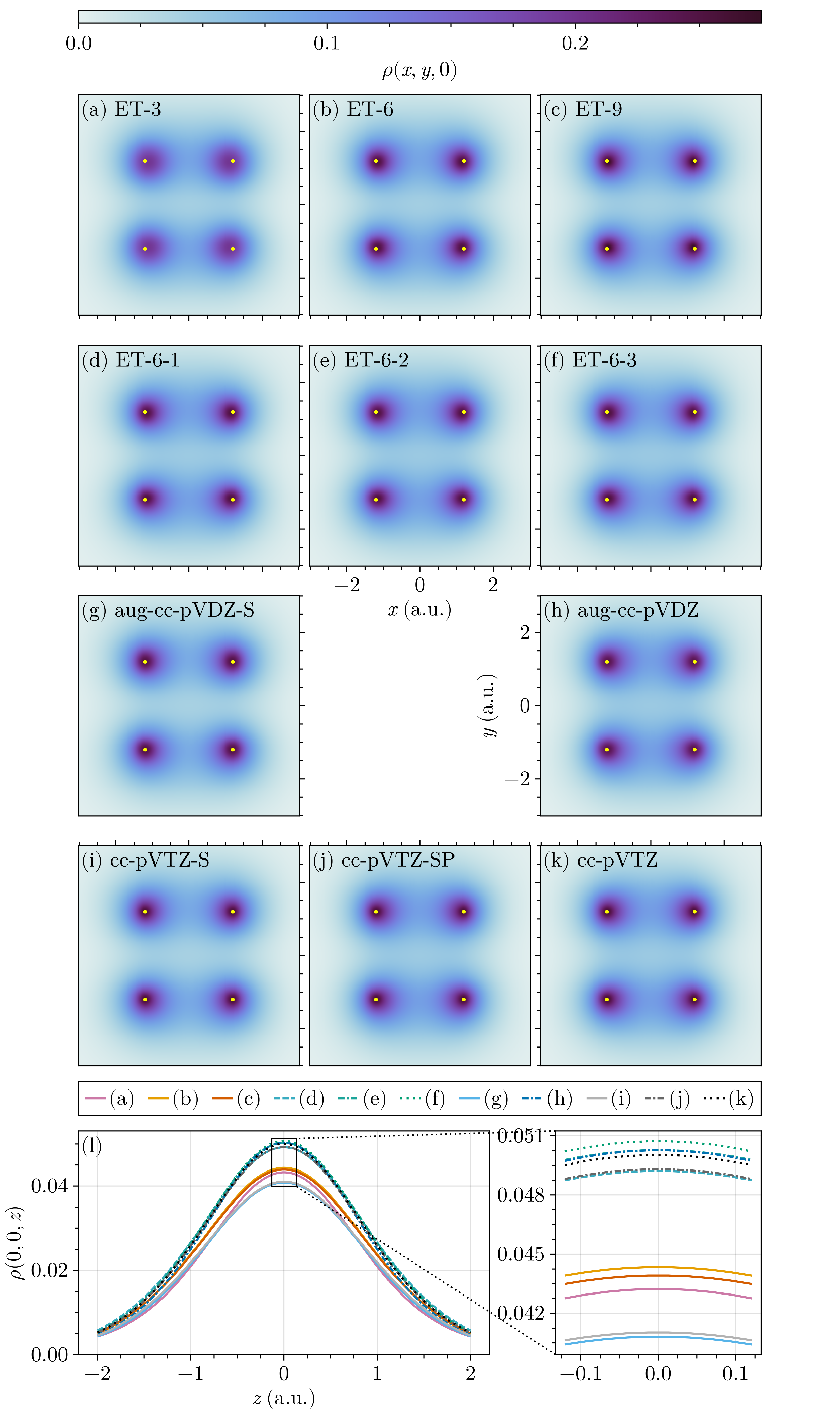}
    \vspace{-15pt}
    \captionof{figure}{Square planar ($2.4$ a.u.).}
    \label{fig:den_H4sqrt2p4}
  \end{minipage}
\end{figure}

\begin{figure}[thb]
  \vspace{-10pt}
  \centering
  \begin{minipage}[t]{0.48\linewidth}
    \centering
    \includegraphics[width=\linewidth]{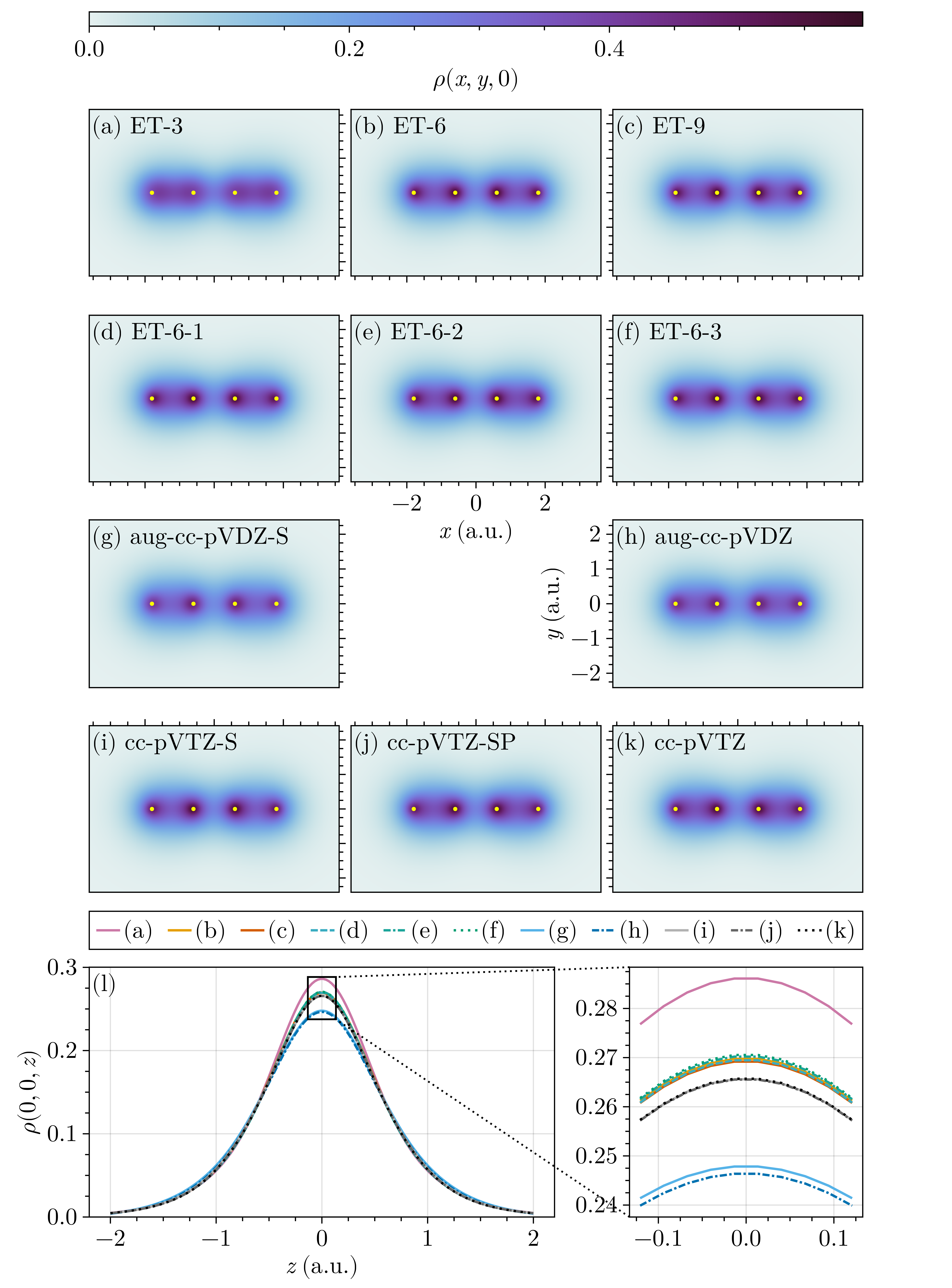}
    \vspace{-15pt}
    \captionof{figure}{Linear chain ($1.2$ a.u.).}
    \label{fig:den_H4_1p2}
  \end{minipage}\hfill
  \begin{minipage}[t]{0.48\linewidth}
    \centering
    \includegraphics[width=\linewidth]{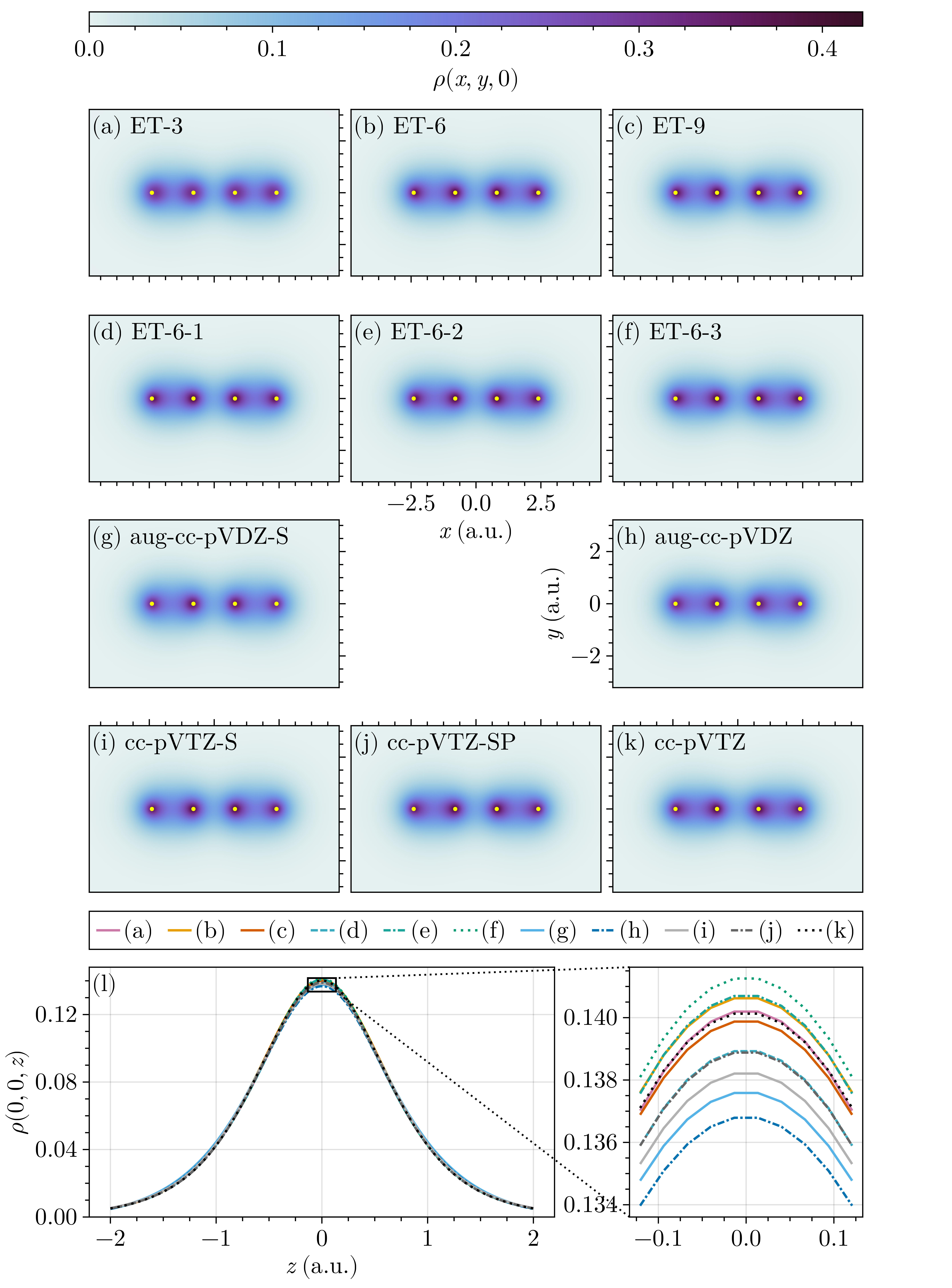}
    \vspace{-15pt}
    \captionof{figure}{Linear chain ($1.6$ a.u.).}
    \label{fig:den_H4_1p6}
  \end{minipage}
  \vspace{-85pt}
\end{figure}

\begin{figure}[thb]
  \vspace{-85pt}
  \centering
  \begin{minipage}[t]{0.48\linewidth}
    \centering
    \includegraphics[width=\linewidth]{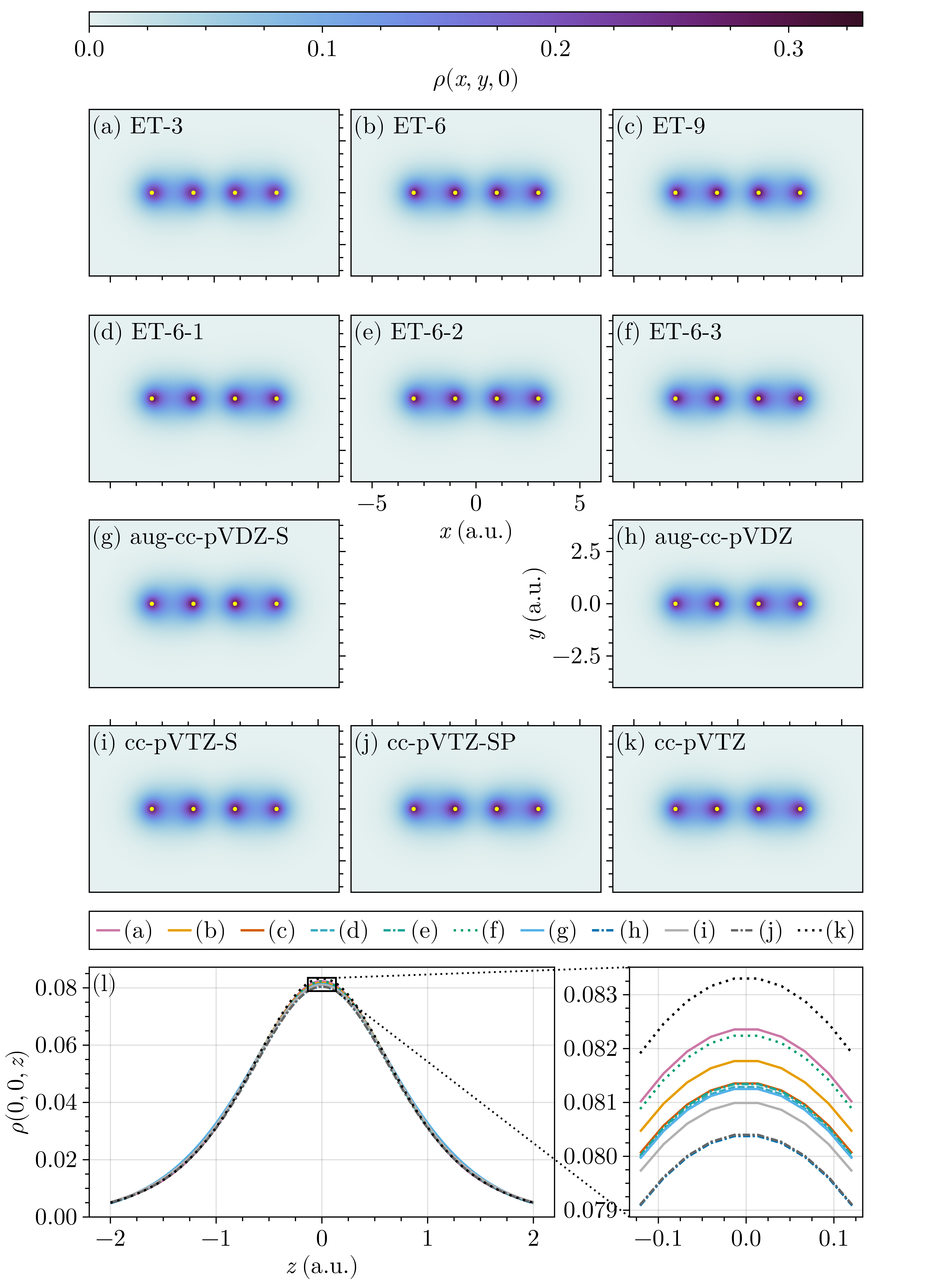}
    \vspace{-15pt}
    \captionof{figure}{Linear chain ($2.0$ a.u.).}
    \label{fig:den_H4_2p0}
  \end{minipage}\hfill
  \begin{minipage}[t]{0.48\linewidth}
    \centering
    \includegraphics[width=\linewidth]{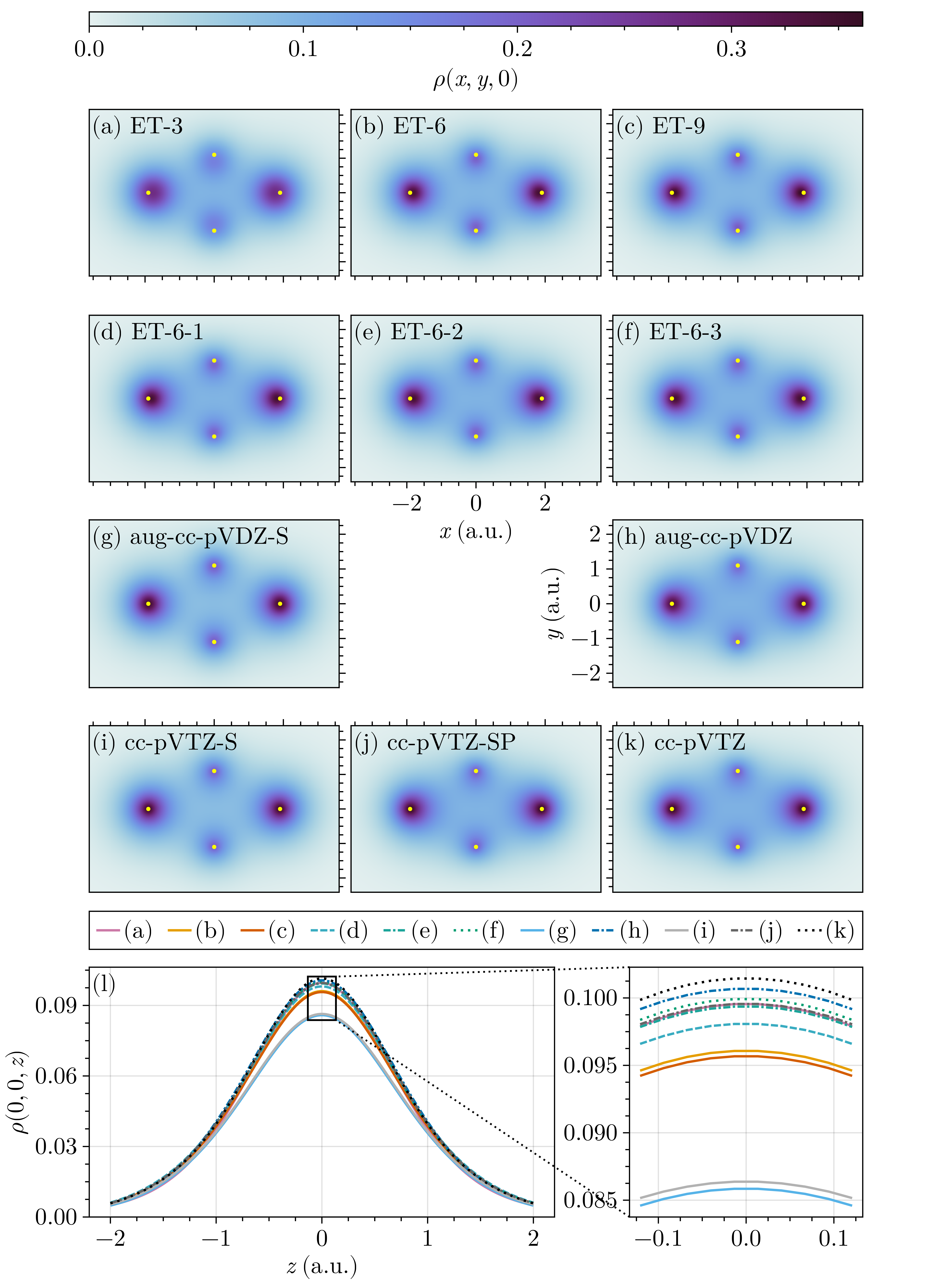}
    \vspace{-15pt}
    \captionof{figure}{Rhombus ($60^{\circ}$, $2.2$ a.u.).}
    \label{fig:den_H4rbs2p2}
  \end{minipage}
  \vspace{-125pt}
\end{figure}

This appendix includes the figures (FIGs.~\ref{fig:den_H4sqrt2p0}--\ref{fig:den_H4rbs2p2}) of HF charge densities for all tested H$_4$ geometries in Sec.~\ref{sec:H4}, computed with the basis sets (and subsets) compared in TABLE~\ref{tab:H4energy}. Within each figure, the panels (i.e., the sub-figures) appear in the same order as the basis sets listed in TABLE~\ref{tab:H4energy}, and each panel is labeled by the corresponding basis abbreviation. For example, “ET-6-1” in FIG.~\ref{fig:den_H4sqrt2p0} denotes the HF charge density obtained with the variational even-tempered basis set having basis–degree pair (6, 1). In each panel, the nuclear positions of the system are marked by the yellow dots.


\end{document}